**Title**

# Testing the Power-Law Hypothesis of the Inter-Conflict Interval

**Authors**

Hiroshi Okamoto,[1]* Iku Yoshimoto,[2] Sota Kato,[3] Budrul Ahsan,[3] Shuji Shinohara[4]

**Affiliations**

1. Department of Bioengineering, School of Engineering, The University of Tokyo, Tokyo, Japan
2. Department of Advanced Social and International Studies, Graduate School of Arts and Sciences, The University of Tokyo, Tokyo, Japan
3. The Tokyo Foundation for Policy Research, Tokyo, Japan
4. School of Science and Engineering, Tokyo Denki University, Saitama, Japan

*Corresponding author. E-mail: okamoto@coi.t.u-tokyo.ac.jp

**Abstract**

The severity of war, measured by battle deaths, follows a power-law distribution. Here, we demonstrate that power law also holds in the temporal aspects of interstate conflicts. A critical quantity is the inter-conflict interval (ICI), the interval between the end of a conflict in a dyad and the start of the subsequent conflict in the same dyad. Using elaborate statistical tests, we confirmed that the ICI samples compiled from the history of interstate conflicts from 1816 to 2014 followed a power-law distribution. We propose an information-theoretic model to account for the power-law properties of ICIs. The model predicts that a series of ICIs in each dyad is independently generated from an identical power-law distribution. This was confirmed by statistical examination of the autocorrelation of the ICI series. Our findings help us understand the nature of wars between normal states, the significance of which has increased since the Russian invasion of Ukraine in 2022.

**Teaser**

The timing of the conflict obeys the power law, the mechanism of which can be explained by hypothesizing that the use of force is a means of interstate communication.

**Introduction**

Investigating the statistical patterns behind historical cases of interstate conflicts is pivotal, especially from the following perspectives: First, if identified, these patterns encourage an inductive approach to conflict mechanisms. This approach, used to formulate hypotheses based on experimental or empirical observations and to test their predictions, has been reliably employed in the natural sciences. In international relations theory, in contrast, the theory-based approach, which attempts to derive a novel theory from plausible assumptions by deduction, has been more favored for solving, for instance, *war's inefficiency puzzle* (Fearon 1995). Second, robust statistical patterns are a prerequisite for estimating the probability of the future occurrence of an interstate conflict in a specified dyad (a pair of states) (Herge et al. 2017). Forecasting the future occurrence of an interstate conflict is beneficial for scholarly understanding of conflict processes and supporting policymaking by international organizations to prevent, manage, or resolve armed conflict, or by individual states to establish national security. From this practical perspective, finding robust statistical patterns in any aspect of interstate conflict is desirable.

In fact, the prediction of event occurrences, such as war, alliance formation, and revolutions, is at the heart of international relations theory, and scholars in the discipline have realized that a proper arrangement of time is required to achieve sophisticated predictions (Beck et al. 1998; Crescenzi & Enterline 2001). Whether we should explicitly theorize the relationship between time and the events under study (Crescenzi & Enterline 2001; Carter & Signorino 2010) or not (Beck 2010), it is now standard practice to account for time in modeling the occurrence of events by including splines or polynomials (Beck et al. 1998; Carter & Signorino 2010) or by using Cox duration models (Metzger & Jones 2022). These attempts suggest that time plays a vital role in international events, the most notable of which is war. Therefore, exploring the temporal structures of conflict processes is a prerequisite to understanding and predicting wars.

Notably, scholars of complexity science, instead of those of mainstream international relations theory, have enthusiastically sought the statistical patterns of conflict. Outstanding findings were made by the English physicist Lewis Fry Richardson more than three-quarters of a century ago (Richardson 1948; Richardson 1960). He found that the severity of war, measured by battle deaths, followed the power law. These findings were later confirmed in more detailed studies (Bohorquez et al. 2009; Cederman et al. 2011; Friedman 2015; Cirrilo & Taleb 2016; Gonzalez 2016; Clauset 2018; Spagat et al. 2018; Cunen et al. 2020). The power-law distribution of war sizes, characterized as fat-tailed, implies the possible occurrence of *black swan* events, such as World War I (WWI) or World War II (WWII). It has also been shown that the severity of other forms of human violence, such as civil war, insurgency, or terrorist attacks, follows the power law (Clauset et al. 2007; Bohorquez et al. 2009; Clauset & Gleditsch 2012; Johnson et al. 2013). Finding statistical patterns in the severity of war and other human violence has inspired the exploration of the mechanism for the escalation of violence, typically attributed to 'critical phenomena' resulting from the operation of positive feedback loops (Cederman, 2003; Bohorquez et al. 2009; Cederman et al. 2011; Johnson et al. 2011; Clauset & Gleditsch 2012; Johnson et al. 2013; DiVita 2020; Johnson-Restrepo et al. 2020). This robust statistical pattern is also used to infer the actual number of casualties of inadequately recorded wars or to examine the risk of the future occurrence of huge wars such as WWI or WWII (Scharpf et al. 2014; Friedman 2015; Cirrilo & Taleb 2016; Clauset 2018; Cunen et al. 2020).

This study shows that the power-law property also resides in the temporal aspects of interstate conflicts. This study focuses on the temporal structure of conflict occurrence rather than the spatial structure, such as the scale of war, because the temporal structure is thought to be more fundamentally related to decision-making by states regarding military action, which is the core process of conflict occurrence. Additionally, robust statistical patterns in the temporal structure can be used more directly to predict the future occurrence of conflict because prediction is a task along the temporal dimension. As a key quantity in our analysis, we define the inter-conflict interval (ICI) as the interval between the end of a conflict in a dyad and the start of the subsequent conflict in the same dyad. ICI samples eligible for statistical analysis were obtained from a dataset provided by the Correlates of War Project. Using an elaborate statistical method to test the power-law hypothesis, we confirmed that the ICIs follow a power-law distribution. We formulated a hypothetical model that accounted for the power-law distribution of ICIs under minimal assumptions. We then tested the prediction of this model, which states that the power law holds true for individual dyads.

# Results

### Terminology: *interstate war*, *militarized interstate dispute*, and *interstate conflict*

First, we specify the definitions of interstate wars, militarized interstate disputes, and interstate conflicts. Precise definitions of the first two terms are provided by the Correlates of War (COW) Project (https://correlatesofwar.org/). An *interstate war* is a series of sustained battles between the armed forces of two or more states that have resulted in at least 1,000 battle deaths (Maoz et al. 2019). A *militarized interstate dispute* is a historical case of conflict in which the threat, display, or use of military force *short of* interstate war by one state is explicitly directed towards the government, official representatives, official forces, property, or territory of another state (Jones et al. 1996; Maoz et al. 2019). In this study, we use the third term, *interstate conflict*, to express the union of militarized interstate disputes and interstate wars.

In a militarized interstate dispute or an interstate war, military action taken by one or both states is often preceded by political issues between them, such as conflicting national interests or disagreements over foreign policy. The inclusion of militarized interstate disputes and interstate wars in our analysis is appropriate if we stand on the view that the use of force, in any form, should be a way to resolve international issues, whereas previous studies on the severity of interstate conflict have focused only on interstate wars (Cederman et al. 2011; Cirrilo & Taleb 2016; Clauset 2018). Our view resonates with the famous thesis of Prussian general and war philosopher Carl von Clausewitz (Howard 2002):

> "WAR IS MERELY THE CONTINUATION OF POLICY BY OTHER MEANS" (Clausewitz 1832)

He derived a corollary from this thesis, arguing that:

> "The political object---the original motive of the war---will thus determine both the military objective to be reached and the amount of effort it requires." (Clausewitz 1832)

The use of force in an actual war must be proportional to the political objectives. Thus, Clausewitz's arguments motivated us to address militarized interstate disputes (MIDs) and interstate wars without distinction.

### Dataset of interstate conflict

We used Dyadic MID Version 4.02 (MID4.02), a dataset provided by the COW Project (Maoz et al 2019). The dataset records interstate conflicts between 1816 and 2014. Each interstate conflict is specified by a dyad (a pair of states) engaged in the conflict and the start and end dates of the conflict.

### Inter-conflict interval

We sought to identify robust statistical patterns behind the temporal structure of the occurrence of interstate conflicts. The inter-conflict interval (ICI) is the critical quantity to this end and is defined as the interval between the end of a conflict in a dyad and the start of the next conflict in the same dyad (**Fig. 1**). We obtained 2,369 ICI samples from MID4.02, each measured in days. These ICI samples were collected from all dyads.

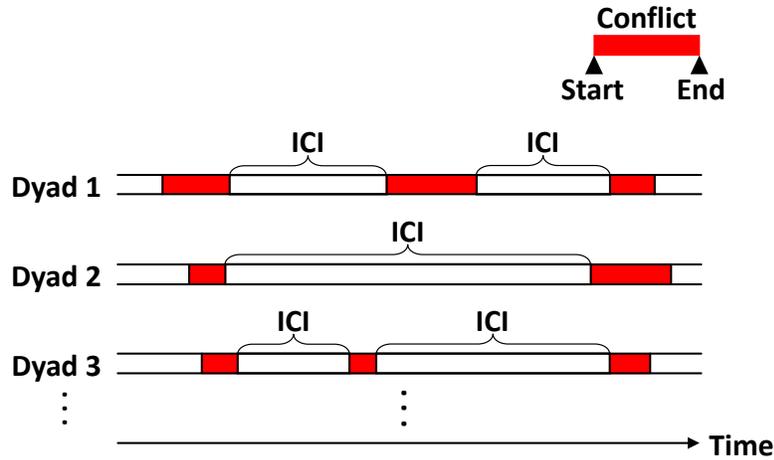

**Figure 1:** Inter-conflict intervals (ICIs). The ICI is the interval between the end of a conflict in a dyad and the start of the next conflict in the same dyad. Each conflict is indicated by the red rectangle.

**Testing the power-law hypothesis of ICIs**

First, the distribution of these 2,369 ICIs binned with a width of 365.25 days (~one year) was examined in log-log and linear-log plots. Falling into a straight line in a log-log or linear-log plot is characteristic of a power-law or exponential distribution, respectively. The linear regression results in both plots suggested that the ICIs followed a power-law distribution ($R^2 = 0.9179$, **Fig. 2a**) instead of an exponential distribution ($R^2 = 0.6905$, **Fig. 2b**).

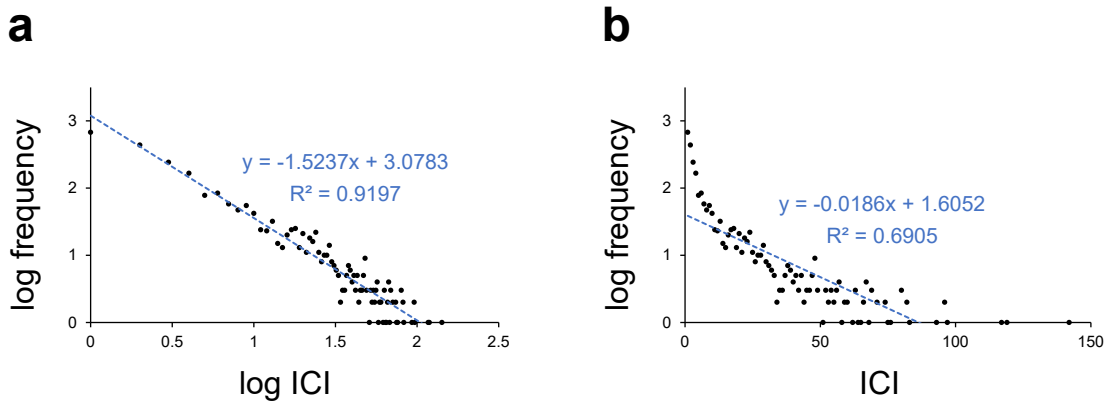

**Figure 2**: Distribution of 2,369 ICI samples collected from all dyads is shown in log-log (**a**) and linear-log (**b**) plots. The bin width for the distribution was chosen as 365.25 days (~one year). The dashed blue line in each panel indicates the linear regression results.

We also examined the power-law fitting using a more elaborate statistical test proposed by Clauset et al. (2009), which we call the Clauset-Shalizi-Neuman (CSN) test. The empirical distributions of the ICI samples would have an artificial upper bound because of the limitations of the recording period, even if they were generable from power-law distributions with infinitely extended tales (see **Materials and Methods** for further details). To consider the possible existence of artificial upper bounds, we used a modified version of the CSN test (mCSN test).

The power-law hypothesis to be examined using the mCSN test is mathematically expressed as follows: $p(x) = x^{-\gamma}/Z(\gamma)$ ($x_{\min} \leq x \leq x_{\max}$). Here, the argument $x$ takes an integer value; $\gamma$ is the exponent of the power-law distribution; $x_{\min}$ and $x_{\max}$ are the lower and upper bounds of the range where the power law holds, respectively; $Z(\gamma) = \sum_{x=x_{\min}}^{x_{\max}} x^{-\gamma}$ is the normalization constant equal to the partition function. The details of the mCSN test are provided in the **Materials and Methods**. In brief, we first estimated the exponent $\gamma$ and lower bound $x_{\min}$, and then calculated the $p$-value. Let $\hat{\gamma}$ and $\hat{x}_{\min}$ be the estimated values of $\gamma$ and $x_{\min}$, respectively. The upper bound $x_{\max}$ was used as the control parameter. Therefore, $\hat{\gamma}$ and $\hat{x}_{\min}$ as well as the $p$-value were given as a function of $x_{\max}$. Clauset et al. (2009) proposed the conservative decision criteria: If $p \leq 0.1$, the power-law hypothesis is ruled out; otherwise, it is plausible. The same criteria were used in this study.

The results of the mCSN tests are shown in **Fig. 3**. The $p$-value exceeded the criteria of 0.1 (indicated by the horizontal dashed line in **Fig. 3a**) for up to $x_{\max}$ slightly longer than 20,000 days (~55 years) (**Fig. 3a**). In **Figs. 3b** and **c**, we observe that $\hat{\gamma}$ and $\hat{x}_{\min}$ are almost constant with $x_{\max}$; $\hat{\gamma}$ is approximately 1.3 and $\hat{x}_{\min}$ is approximately 250 days (<1 year). From these observations, we conclude that the ICI obeys the power law for the range of 250–20,000 days. Approximately 80% of the ICI samples were within this range for $x_{\max} = 20{,}000$ (**Fig. 3d**).

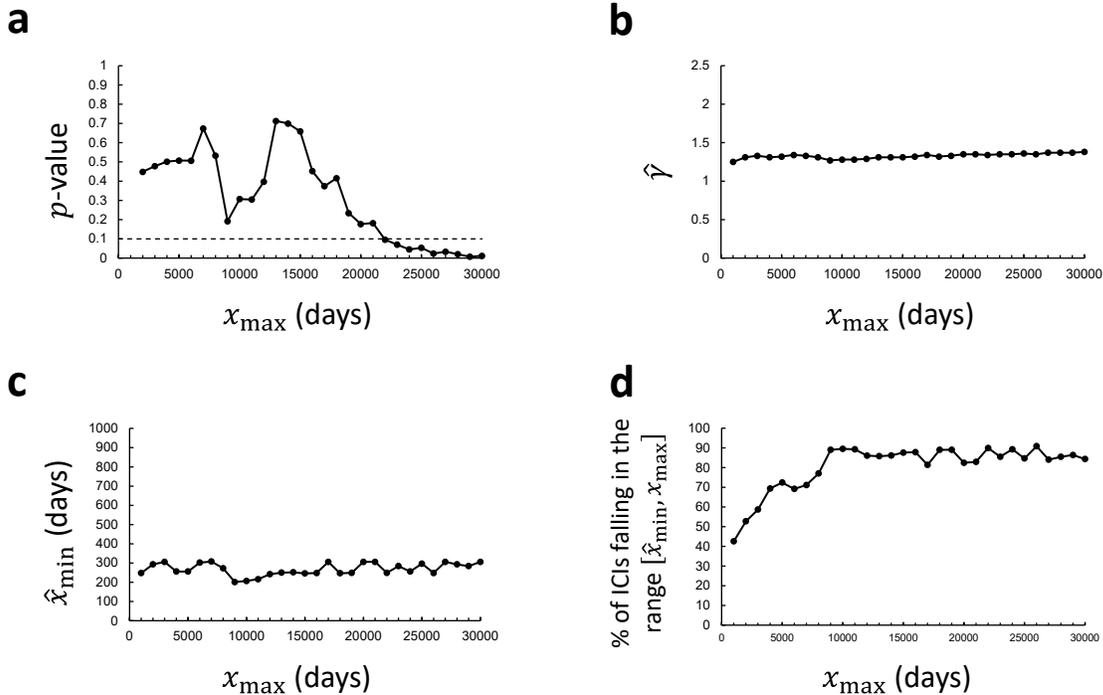

**Figure 3**: Results of the mCSN test applied to 2,369 ICI samples collected from all dyads. This test reveals the plausibility of the power-law hypothesis expressed in the form: $p(x) \propto x^{-\gamma}$ for $x_{\min} \leq x \leq x_{\max}$, where $x_{\min}$ and $x_{\max}$ are the lower and upper bounds of the domain in which the power law holds, respectively. The upper bound $x_{\max}$ is treated as a control parameter, and the optimal values of $\gamma$ and $x_{\min}$ are estimated for each value of $x_{\max}$. (**a**) The $p$-value of the mCSN test is plotted as a function of $x_{\max}$. The horizontal dashed line indicates the criteria of 0.1, for the $p$-value above which the power-law hypothesis is plausible. (**b**) The estimated power-law exponent $\hat{\gamma}$ is plotted as a function of $x_{\max}$. (**c**) The estimated lower bound $\hat{x}_{\min}$ is plotted as a function of $x_{\max}$. (**d**) The ratio of ICIs (out of the total, 2,369) that fall in the power-law holding domain ($\hat{x}_{\min} \leq x \leq x_{\max}$) is plotted as a function of $x_{\max}$.

## Information-theoretic model of interstate conflict

Next, we built a hypothetical model that accounted for the observed power-law properties of ICIs. Consider a dyad of states A and B. Suppose that the $n$-th conflict $C_n$ is provoked by either state. We refer to the state that triggers conflict as the provoker and the opponent state as the withstander. Conflict $C_n$ is characterized by the time of its occurrence and the military actions taken during the conflict. Let this time and the military actions be represented by stochastic variables $T_n$ and $\mathbf{X}_n$, respectively. For simplicity, we assume that the period bounded by the start and end of a conflict contracts to a point. Therefore, $T_n$ takes the real value $t_n \in R^1$. In contrast, corresponding to the various possibilities of the course of a war, $\mathbf{X}_n$ would take multidimensional values $\mathbf{x}_n$ that would be categorical or numerical. Furthermore, each military action may be led by either the provoker or the withstander, as contingent switching between offense and defiance is the case during the course of war. Nevertheless, in the following discussion, we formally deal with $\mathbf{X}_n$ without addressing its mathematical details.

After the settlement of conflict $C_n$, a post-conflict order is established, whether or not it is what the provoker desires. Then, either state, which is discontent with the status of this order and wants to change it to what is more favorable to it, intends to provoke the next conflict $C_{n+1}$. The provoker of conflict $C_{n+1}$ may or may not be the same as that of conflict $C_n$. The time of conflict $C_{n+1}$ and military actions taken during this conflict are represented by the stochastic variables $T_{n+1}$ and $\mathbf{X}_{n+1}$, respectively.

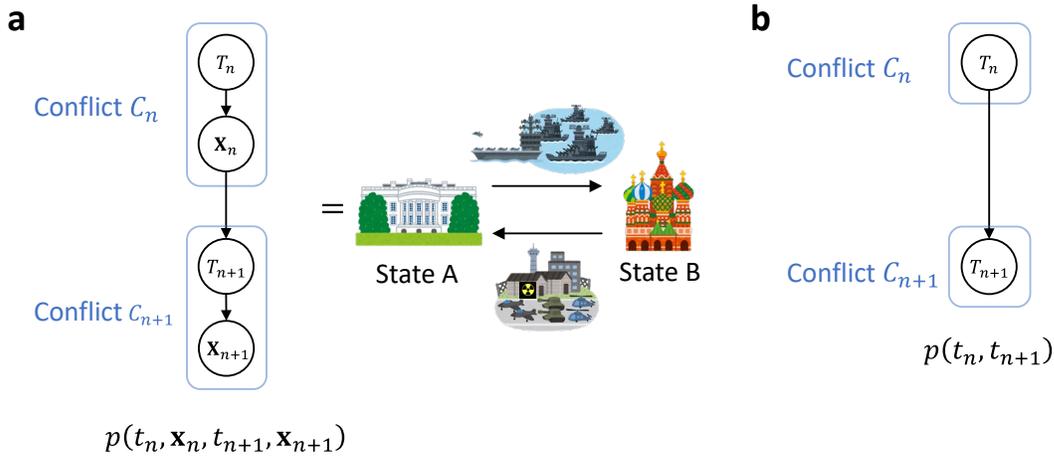

**Figure 4**: Graphical models describing the causal relations between stochastic variables representing consecutive occurrences of conflicts $C_n$ and $C_{n+1}$. Stochastic variables $T_n$ and $\mathbf{X}_n$ represent the time of occurrence of conflict $C_n$ and the military operations taken during the course of this conflict, respectively. (**a**) A graphical model representing the causal relations between $T_n$, $\mathbf{X}_n$, $T_{n+1}$, and $\mathbf{X}_{n+1}$ is shown on the left side, which corresponds to the joint probability $p(t_n, \mathbf{x}_n, t_{n+1}, \mathbf{x}_{n+1})$. The illustration on the right-

hand side describes our hypothesis that the amount of information transferred from $\{T_n, \mathbf{X}_n\}$ to $\{T_{n+1}, \mathbf{X}_{n+1}\}$ is equivalent to the amount of information mutually exchanged between the two states through their engagement in consecutive conflicts $C_n$ and $C_{n+1}$. (**b**) A graphical model representing the causal relation between $T_n$ and $T_{n+1}$, which is obtained by marginalizing the graphical model on the left side of (**a**) over $\mathbf{X}_n$ and $\mathbf{X}_{n+1}$ and corresponds to the probability $p(t_n, t_{n+1})$.

The end (purpose) of war is to attain a political objective, and military action is a means to achieve this objective. Both the provoker and the withstander conceived their own purposes. The provoker's purpose is to compel the other to submit to its will, whereas the withstander's purpose is to compel the provoker to withdraw. The variable $T_n$ describes when a political disagreement between the two states becomes critical and either or both states decide to resolve this by force. In this respect, $T_n$ reflects the purpose of the war. In fact, $T_n$ encodes when the purpose is conceived but does not say what it is. Because the means should be aligned with the purpose, $\mathbf{X}_n$ instead of $T_n$ reflect what the purpose is. As conceiving a purpose precedes choosing the means, $T_n$ causally precedes $\mathbf{X}_n$. In summary, the causal relationships between $T_n$, $\mathbf{X}_n$, $T_{n+1}$, and $\mathbf{X}_{n+1}$ are expressed by the graphical model shown on the left side of **Fig. 4a**, which corresponds to the joint probability $p(t_n, \mathbf{x}_n, t_{n+1}, \mathbf{x}_{n+1})$.

According to information theory, the amount of information carried by stochastic variables $\{T_n, \mathbf{X}_n\}$ is measured by entropy:

$$H[T_n, \mathbf{X}_n] = -\int dt_n d\mathbf{x}_n\ p(t_n, \mathbf{x}_n) \log p(t_n, \mathbf{x}_n). \tag{1}$$

The amount of information successfully received by the stochastic variables $\{T_{n+1}, \mathbf{X}_{n+1}\}$ out of the total sent by $\{T_n, \mathbf{X}_n\}$, which is precisely the entropy $H[T_n, \mathbf{X}_n]$, is measured by mutual information:

$$\begin{aligned} &I\big[\{T_n, \mathbf{X}_n\}, \{T_{n+1}, \mathbf{X}_{n+1}\}\big] \\ &= \int dt_n d\mathbf{x}_n dt_{n+1} d\mathbf{x}_{n+1} p\big(\{t_n, \mathbf{x}_n\}, \{t_{n+1}, \mathbf{x}_{n+1}\}\big) \log \frac{p\big(\{t_n, \mathbf{x}_n\}, \{t_{n+1}, \mathbf{x}_{n+1}\}\big)}{p(t_n, \mathbf{x}_n) p(t_{n+1}, \mathbf{x}_{n+1})}. \end{aligned} \tag{2}$$

It is reasonable to hypothesize that the amount of information transferred from $\{T_n, \mathbf{X}_n\}$ to $\{T_{n+1}, \mathbf{X}_{n+1}\}$ corresponds to the amount of information mutually exchanged between the two states through their engagement in consecutive conflicts $C_n$ and $C_{n+1}$ (see the illustration on the right-hand side of **Fig. 4a**).

Equation (2) can be arranged as

$$I\big[\{T_n, \mathbf{X}_n\}, \{T_{n+1}, \mathbf{X}_{n+1}\}\big] = I[T_n, T_{n+1}] + I[\mathbf{X}_n, \mathbf{X}_{n+1} | T_n, T_{n+1}], \tag{3}$$

where

$$\begin{aligned} I[T_n, T_{n+1}] &= H[T_n] + H[T_{n+1}] - H[T_n, T_{n+1}], \\ I[\mathbf{X}_n, \mathbf{X}_{n+1} | T_n, T_{n+1}] &= H[\mathbf{X}_n | T_n] + H[\mathbf{X}_{n+1} | T_{n+1}] - H[\mathbf{X}_n, \mathbf{X}_{n+1} | T_n, T_{n+1}]. \end{aligned}$$

Thus, the total amount of information exchanged between the two states through their engagement in consecutive conflicts $C_n$ and $C_{n+1}$ is equivalent to the sum of the mutual information $I[T_n, T_{n+1}]$ and $I[\mathbf{X}_n, \mathbf{X}_{n+1}|T_n, T_{n+1}]$. We interpret $I[T_n, T_{n+1}]$ as the amount of information exchanged at the national strategy level, and $I[\mathbf{X}_n, \mathbf{X}_{n+1}|T_n, T_{n+1}]$ as that exchanged at the military operation level. The latter is particularly relevant to the extent to which battle lessons from military operations taken during conflict $C_n$ influence those taken during conflict $C_{n+1}$. We considered conflicts to be event units, and the success or failure of military operations conducted during each conflict was outside the scope of this study. That is, our main interest is the communication between the two states at the national strategy level. Therefore, we focus on $I[T_n, T_{n+1}]$. In doing so, we marginalize the graphical model in **Fig. 4a** over $\mathbf{X}_n$ and $\mathbf{X}_{n+1}$ to obtain the graphical model in **Fig. 4b**, which corresponds to the probability $p(t_n, t_{n+1}) = \int d\mathbf{x}_n d\mathbf{x}_{n+1} p(t_n, \mathbf{x}_n, t_{n+1}, \mathbf{x}_{n+1})$.

Thus, our consideration leads to the intriguing notion that the amount of information exchanged between the two states at the national strategy level depends only on the relative timing of their engagement in consecutive conflicts. In the present study, we followed this notion without further verification. Future studies should address historical cases of interstate conflicts to verify this notion empirically.

The interval between conflicts $C_n$ and $C_{n+1}$, now given by $T_{n+1} - T_n$, also served as a stochastic variable. Once $p(t_n, t_{n+1})$, the joint probability of $T_n$ and $T_{n+1}$, is known, $p(t_{n+1} - t_n)$, the distribution of $T_{n+1} - T_n$, can be easily calculated. Therefore, we determined the functional forms of $p(t_n, t_{n+1})$. Information theory states that a probability distribution that actually exists maximizes entropy. In general, entropy maximization is performed under constraints that specify the objects or phenomena of interest. To define the constraints in our case, we assume that states A and B, struggling with their national interests and survival, will behave according to the trade-off between the principle of promptness and the principle of seriousness.

The need for the first principle of promptness can be easily understood. Suppose that the status quo is unfavorable for state A. The longer this status continues, the more state A will incur losses in national interest. To prevent further losses, state A intends to take military action in any form against state B to change the status quo as promptly as possible. The principle of promptness implies a behavioral tendency to avoid wasting time.

The second principle, seriousness, is closely related to communication in an information-theory sense. Remind Clausewitz's fundamental thesis: "War is merely the continuation of policy by other means." We now interpret this thesis from the perspective of modern information theory, rephrasing it as follows. The use of military force is a means of interstate communication. To formulate interstate communication through force within the framework of information theory, it is useful to note Clausewitz's argument.

> "War is no pastime; it is no mere joy in daring and winning, no place for irresponsible enthusiasts. It is a serious means to a serious end, …" (Clausewitz 1832).

This implies that the state responds seriously to an opponent's move. (Serious responses do not necessarily mean rational responses; see **Discussion**). From an information-theoretic perspective, a pair of states acts in such a way that the communication between variables $\{T_n, \mathbf{X}_n\}$ and $\{T_{n+1}, \mathbf{X}_{n+1}\}$ (**Fig. 4a**) is as efficient as possible. Even after marginalization (**Fig. 4b**), the remaining variables $T_n$ and $T_{n+1}$ should be as mutually dependent as possible. The principle of seriousness implies that there is no room for behavioral redundancy in the theater, where rival states act per their national interests and survival.

To achieve 'a serious means to a serious end,' the principle of promptness alone is inadequate. Suppose that conflict occurs at a high frequency, following this principle; however, the timing of each conflict occurrence is statistically independent of that before it (this is the case if a conflict occurs following a Poisson process). This implies that conflict occurs only erratically, which is the opposite of seriousness.

The constraints for entropy maximization to determine the functional forms of $p(t_n, t_{n+1})$ are defined by the principles above. For mathematical simplicity, we consider the case where $T_n$ and $T_{n+1}$ take continuous values: $-\infty < t_n < x + \Delta \leq t_{n+1} < +\infty$, where $\Delta \,(> 0)$ is the minimum length of ICI. The constraint representing the principle of promptness is defined as the force required to reduce $T_{n+1} - T_n$. Because $T_{n+1} - T_n$ is a stochastic variable, its statistical mean is reduced, not its raw value. There are several types of statistical methods, such as arithmetic or geometric. Therefore, the question arises: What kind of statistical means should we choose? More specifically, what kinds of statistical means of $T_{n+1} - T_n$ do states behave to reduce? We leave aside this problem and instead consider the generalized mean, which can express a variety of statistical means by varying the parameterization. We later demonstrate that the parameterization is determined by the second principle.

The generalized mean of $T_{n+1} - T_n$ is given by

$$\mathrm{E}^{(\mathrm{gen})}_{T_n, T_{n+1}}[T_{n+1} - T_n] = \left( \int_{-\infty}^{+\infty} dt_n \int_{t_n + \Delta}^{+\infty} dt_{n+1} \, p(t_n, t_{n+1}) (t_{n+1} - t_n)^m \right)^{1/m}, \tag{4}$$

where $m$ is the parameter characterizing the generalized mean and $\Delta \,(> 0)$ is the minimum length of the possible interval between conflicts $C_n$ and $C_{n+1}$. By varying $m$, Eq. (4) yields various statistical methods. For example, Eq. (4) is equal to the arithmetic mean for $m = 1$ and approaches the geometric mean for $m \to 0$.

The joint entropy of $T_{n+1}$ and $T_n$ is hence given by

$$S[T_n, T_{n+1}] = -\int_{-\infty}^{+\infty} dt_n \int_{t_n + \Delta}^{+\infty} dt_{n+1} \, p(t_n, t_{n+1}) \log p(t_n, t_{n+1})$$

$$- \gamma \log \mathrm{E}^{(\mathrm{gen})}_{T_n, T_{n+1}}[T_{n+1} - T_n] - \lambda \left( \int_{-\infty}^{+\infty} dt_n \int_{t_n + \Delta}^{+\infty} dt_{n+1} \, p(t_n, t_{n+1}) - 1 \right). \tag{5}$$

The first term on the right-hand side represents the Shannon's entropy. The second term is introduced according to the first principle of promptness and expresses the force required to reduce the generalized mean of $T_{n+1} - T_n$; the coefficient $\gamma \,(> 0)$ controls the strength of this force. The third term, where $\lambda$ is a Lagrange multiplier, ensures the normalization condition that $p(t_n, t_{n+1})$ are summed to unity. Maximizing entropy (5) with respect to $p(t_n, t_{n+1})$, with rescaling of $\gamma / \mathrm{E}^{(\mathrm{gen})}_{T_{n+1}, T_n}[T_{n+1}, T_n] \to \gamma$, yields

$$p(t_n, t_{n+1}) = \frac{1}{Z(\gamma, m)} \exp\left[-\frac{\gamma}{m}(t_{n+1} - t_n)^m\right], \tag{6}$$

where

$$Z(\gamma, m) = \int_{-\infty}^{+\infty} dt_n \int_{t_n+\Delta}^{+\infty} dt_{n+1} \exp\left[-\frac{\gamma}{m}(t_{n+1} - t_n)^m\right] \tag{7}$$

is the normalization factor. As expected, Eq. (6) becomes equal to the exponential distribution $p(t_n, t_{n+1}) \propto \exp[-\gamma(t_{n+1} - t_n)]$ for $m = 1$ and approaches the power-law distribution $p(t_n, t_{n+1}) \propto (t_{n+1} - t_n)^{-\gamma}$ for $m \to 0$ (Visser 2013). For $p(t_n, t_{n+1})$ to be normalized, $m$ should be nonzero positive.

Next, we demonstrate that the value of $m$ is determined by the second principle of seriousness. As previously discussed, this principle makes stochastic variables $T_n$ and $T_{n+1}$ mutually dependent as much as possible. Information theory states that the mutual dependence between stochastic variables can be estimated by mutual information:

$$I[T_n, T_{n+1}] = \int_{-\infty}^{+\infty} dt_n \int_{t_n+\Delta}^{+\infty} dt_{n+1}\, p(t_n, t_{n+1}) \log \frac{p(t_n, t_{n+1})}{p(t_n) p(t_{n+1})}, \tag{8}$$

where $p(t_n) = \int_{-\infty}^{+\infty} p(t_n, t_{n+1}) dt_{n+1}$ and $p(t_{n+1}) = \int_{-\infty}^{+\infty} p(t_n, t_{n+1}) dt_n$ are marginal probabilities. Using the forms of Eqs. (6) and (7), and taking $\Delta \to 0$, we can analytically calculate the right-hand side of Eq. (8) to obtain

$$I[T_n, T_{n+1}] = f(m, \gamma) + \text{constant}, \tag{9}$$

$$f(m, \gamma) = \frac{1}{m}\log\gamma - \left(\frac{1}{m} - 1\right)\log m - \log\Gamma\left(\frac{1}{m}\right) - \frac{1}{m}, \tag{10}$$

where $\Gamma(\cdot)$ denotes the gamma function. The principle of seriousness argues that mutual information $I[T_n, T_{n+1}]$ should be maximized. **Fig. 5** shows $f(m, \gamma)$ as functions of $m\ (> 0)$ and $\gamma\ (\geq 1)$. For each value of $\gamma$, $f(m, \gamma)$ is maximized for $m \to +0$. Thus, the principle of seriousness, which is embodied by the maximization of mutual information, leads to the power-law distribution of $\tau \equiv t_{n+1} - t_n$:

$$p(\tau) = \frac{\gamma - 1}{\Delta^{-\gamma+1}} \tau^{-\gamma} \quad (\Delta \leq \tau < +\infty). \tag{11}$$

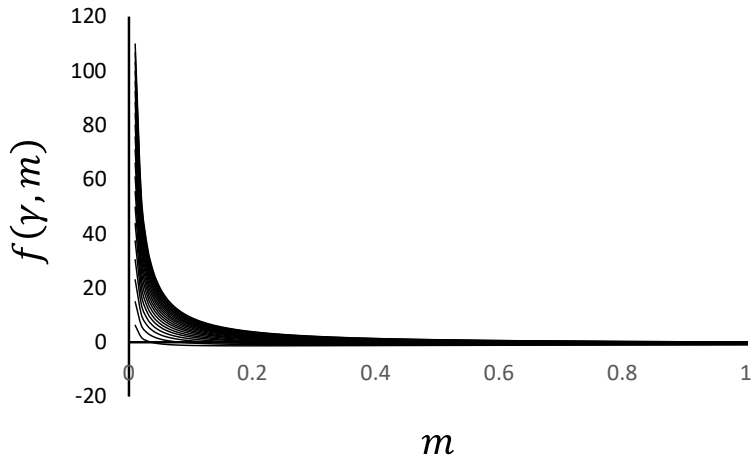

**Fig. 5:** Mutual information $I[T_n, T_{n+1}] = f(\gamma, m) +$ constant as functions of $\gamma$ and $m$. The analytical form of $f(\gamma, m)$ is given by Eq. (10). The curves in the coordinate plane plot $f(\gamma, m)$ as a function of $m$ for different values of $\gamma$ (varied from 1.0 to 3.0 in 0.01 increments, with the lowest curve for $\gamma = 1.0$). For any value of $\gamma$ ($\geq 1$), $f(\gamma, m)$ is maximally extremized at $m \to +0$.

**Dilution of the power-law process: relation between the model and observation**

Because the interval between consecutive occurrences of conflict, but not the duration of each conflict itself, was of interest, we prescribed the duration of each conflict to be contracted to a point in time. With this mathematical simplification, the conflict occurrence in each dyad can be viewed as a point process (**Fig. 6a,** filled black circles). Our information-theoretic model predicts that the point-to-point intervals of this process follow a power-law distribution.

**a**

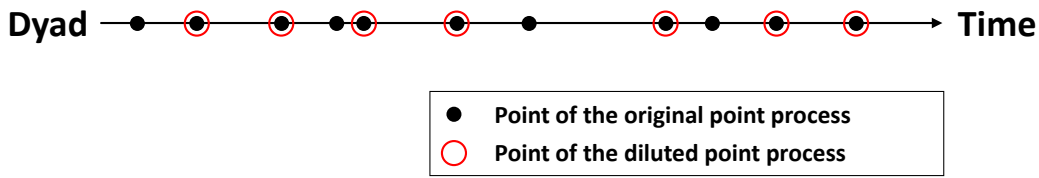

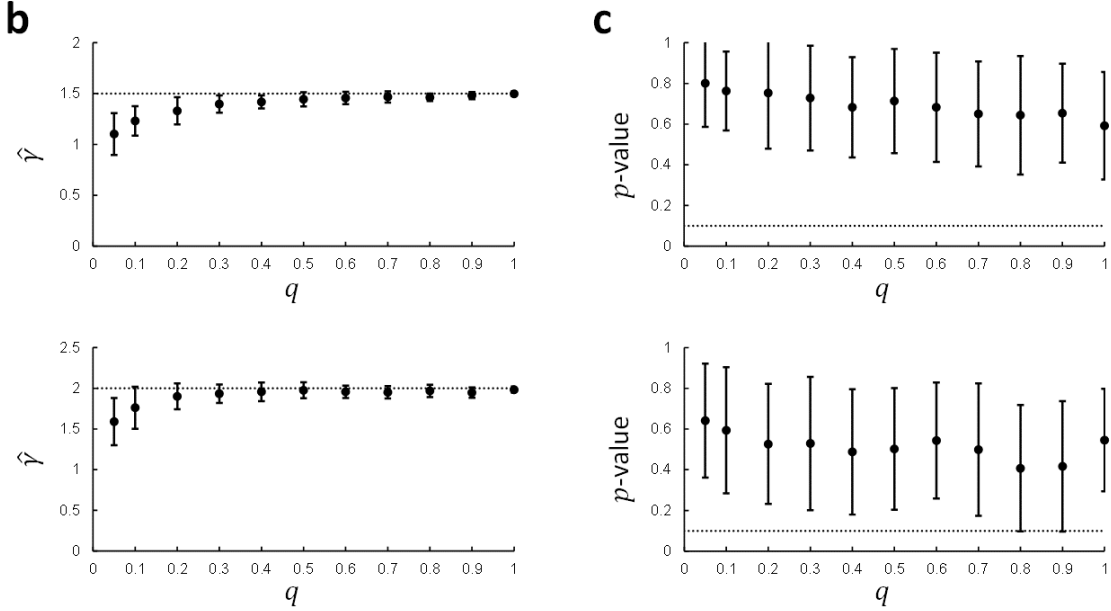

**Figure 6:** (**a**) Illustration of a point process whose point-to-point intervals are supposed to follow a power-law distribution. The chain of filled black circles represents an original point process supposed to follow a power-law distribution. This process is diluted by probabilistically maintaining or discarding each point. The chain of maintained points, indicated by the blank red circles surrounding them, constitutes a diluted point process. The original point process models true occurrences of conflict in the history, regardless of whether they are recorded in the dataset. The diluted point process models conflict occurrences that are actually recorded in the dataset. (**b**) An original point process following a power-law distribution ($p(x) \propto x^{-\gamma}$ for $x = 1, 2, 3, \cdots$) is diluted with the maintaining probability $q$ (hence with the discarding probability $1 - q$). We conducted the mCSN test applied to the point-to-point intervals of 100 processes obtained by the probabilistic dilution. The power-law exponent $\hat{\gamma}$ estimated by this test is plotted as a function of $q$ (upper and lower panels for $\gamma = 1.5$ and $\gamma = 2.0$, respectively). The horizontal dotted lines in both panels indicate the power-law exponent $\gamma$ of the original process. (**c**) The $p$-value of the mCSN test is plotted as a function of $q$ (upper and lower panels for $\gamma = 1.5$ and $\gamma = 2.0$, respectively). The horizontal dotted lines in both panels indicate the criteria of $0.1$, for the $p$-value above which the power-law hypothesis is plausible. In (**b**) and (**c**), the error bars indicate the standard deviations.

However, we should be aware of the possibility of a recording bias. Some cases of militarized interstate disputes may have been overlooked in the collection of data and were not recorded. Therefore, the point process compiled from the dataset is obtained by diluting the original point process generated by the model (**Fig. 6**, blank red circles).

Therefore, it is necessary to examine whether the diluted point process also follows the power law and, if so, whether the power-law exponent for the diluted point process is equal to that for the original point process. The power-law distribution given by Eq. (11) has a lower bound $\Delta$ ($> 0$) in the domain. For $\Delta \to 0$, the diluted point process follows the same power law as the original process, owing to the scale-invariant property of the power-law distribution. However, in reality, $\Delta$ would be slightly greater than zero because a minimum length of time is required (for example, to redeploy resources) before the invocation of the next conflict. To examine whether the point process obtained by diluting the original power-law process with $\Delta > 0$ also follows the power law, we conducted the following numerical experiments: A sample of the point process is generated so that point-to-point intervals follow a power-law distribution with $\Delta = 1$. The generated point process is then diluted with probability $q$; that is, each point is left and abandoned with probabilities $q$ and $1 - q$, respectively. The point-to-point intervals collected from the diluted process then undergo the mCSN test to calculate the $p$-value and estimate the fitted power-law exponent $\hat{\gamma}$.

The experimental results are shown in **Figs. 6b** and **c**. The $p$-value (**Fig. 6c**) and fitted power-law exponent $\hat{\gamma}$ (**Fig. 6b**) are plotted as a function of the probability $q$. For any $q$, the $p$-value averaged over 100 calculations is substantially larger than criteria of 0.1 (**Fig. 6c**), indicating that the power-law hypothesis for the diluted point process is plausible. The fitted power-law exponent decreases from the original value for $\gamma$ as $q$ decreases (**Fig. 6b**). These results demonstrate that point-to-point intervals collected from the diluted process, which model observed ICIs, also follow a power-law distribution, although its exponent is reduced from the original $\gamma$.

**Mixture of power-law distributions**

Our information-theoretic model predicts that conflict occurrences in each dyad follow the power law. In addition, the power-law exponent may differ for each dyad because it originates from a predefined value for $\gamma$, which is not necessarily consistent for every dyad. We will later see that the power-law exponent inferred from the real data varies from dyad to dyad. Therefore, the distribution of ICIs collected from all dyads, which we have shown to follow a power-law distribution (**Fig. 2**), should be a mixture of power-law distributions originating from different dyads, with possibly different exponents. Therefore, verifying whether a mixture of power-law distributions can be approximated accurately using a single power-law distribution is necessary. Indeed, we have accomplished this, the detailed descriptions of which are provided in **Materials and Methods**.

**Testing the power-law hypothesis in individual dyads**

Our information-theoretic model predicts that ICIs collected from individual dyads will follow separate power laws. To test this prediction, we applied the mCSN test to seven dyads: CHN-RUS, CHN-US, GMY-FRN, IND-PAK, IRN-IRQ, ISR-SYR, and RUS-US. We used the following abbreviations: CHN (China), FRN (France), GMY (Germany), IND (India), IRN (Iran), IRQ (Iraq), ISR (Israel), SYR (Syria), and US (the United States). We chose these seven dyads because they provided the number $N \geq 20$ of ICI samples that were likely eligible for statistical examination.

The results of the mCSN tests are presented in **Table 1**. In this test, $x_{\max}$ was chosen as $\max_{n} x_n$, which was the maximum ICI sample for each dyad. Noticeably, the power-law hypothesis of ICIs was plausible for all the dyads we examined ($p > 0.1$ for every dyad). The ratio $N_D/N$, where $N$ and $N_D$ are the total number of ICIs and the number of ICIs equal to or larger than the estimated lower bound $\hat{x}_{\min}$, respectively, was substantially large (>0.7) for every dyad, indicating that the power law holds for a wide range of ICI.

We also compared the power-law hypothesis with the alternative hypothesis that the ICIs follow an exponential distribution. From a set of ICI samples such that $\hat{x}_{\min} \leq \text{ICI} \leq x_{\max}$, 100 pseudo datasets were synthesized using the bootstrap process. We calculated the maximum log-likelihood of the exponential and power-law distributions for each synthesized dataset. A paired $t$-test was conducted to examine whether the maximum log-likelihood of the power-law distribution ($\log L^{(\text{p.l.})}$) was significantly larger than that of the exponential distribution ($\log L^{(\exp)}$). The results summarized in **Table 2** show that the power-law distribution is significantly more plausible than the exponential distribution for every dyad. Thus, we concluded that the ICIs in each dyad followed a power-law distribution, which is consistent with the predictions of our model.

|            | CHN-RUS | CHN-US | FRN-GMY | IND-PAK | IRN-IRQ | ISR-SYR | RUS-US |
|---|---|---|---|---|---|---|---|
| $N$        | 38      | 29     | 20      | 33      | 29      | 33      | 39     |
| $\hat{x}_{\min}$ | 155 | 196 | 46 | 219 | 104 | 153 | 278 |
| $N_D$      | 34      | 26     | 20      | 24      | 27      | 25      | 28     |
| $N_D/N$    | 0.895   | 0.897  | 1.000   | 0.727   | 0.931   | 0.758   | 0.718  |
| $\hat{\gamma}$ | 1.11 | 1.51 | 0.96 | 2.17 | 1.65 | 1.56 | 2.01 |
| $p$-value  | 0.8852* | 0.2151* | 0.9656* | 0.2114* | 0.7105* | 0.9438* | 0.293* |

**Table 1:** Results of the mCSN test of the power-law hypothesis expressed in the form: $p(x) = x^{-\gamma}/Z(\gamma)$ for $x_{\min} \leq x \leq x_{\max}$. Here, the value of $x_{\max}$ is chosen as the maximum length of ICI samples, and the normalization factor is given by $Z(\gamma) = \sum_{x=x_{\min}}^{x_{\max}} x^{-\gamma}$. $N$: the number of ICI samples for each dyad. $\hat{x}_{\min}$: the estimated value of $x_{\min}$. $N_D$: the number of ICI samples within the domain $\hat{x}_{\min} \leq x \leq x_{\max}$. $N_D/N$: the ratio of ICI samples within the domain. $\hat{\gamma}$: the estimated value of the power-law exponent $\gamma$. The bottom row lists the $p$-value of the mCSN test. For the $p$-value larger than the criteria of 0.1, as indicated by the asterisk (*), the power-law hypothesis is plausible.

|            | CHN-RUS | CHN-US | FRN-GMY | IND-PAK | IRN-IRQ | ISR-SYR | RUS-US |
|---|---|---|---|---|---|---|---|
| $\langle \log \hat{L}^{(\text{p.l.})} \rangle_B - \langle \log \hat{L}^{(\exp)} \rangle_B$ | 1.425 | 12.5185 | 5.507 | 1.944 | 7.284 | 3.148 | 9.358 |
| $p$-value  | 4.90E-07 | 3.65E-56 | 3.88E-28 | 8.26E-08 | 1.50E-14 | 2.04E-27 | 5.24E-26 |

**Table 2:** The upper row lists the mean difference $\langle \log \hat{L}^{(\text{p.l.})} \rangle_B - \langle \log \hat{L}^{(\exp)} \rangle_B$ for each dyad. The mean $\langle \log \hat{L}^{(\text{p.l.})} \rangle_B$ was calculated by averaging the loglikelihood for the power-law hypothesis over $B = 100$ pseudo series of ICIs generated using the bootstrap process. The mean $\langle \log \hat{L}^{(\exp)} \rangle_B$ of the loglikelihood for the exponential-distribution hypothesis was calculated similarly. Positive values of the quantity $\langle \log \hat{L}^{(\text{p.l.})} \rangle_B - \langle \log \hat{L}^{(\exp)} \rangle_B$ indicate that the power-law hypothesis is more likely than the exponential-distribution hypothesis. The bottom row lists the $p$-value of the paired $t$-test for each dyad to demonstrate the significance of the positivity or negativity of this quantity.

The estimated power-law exponent $\hat{\gamma}$ varies from dyad to dyad, ranging from ~1.0 to ~2.0 (**Table 1**). These estimated values were robust, as confirmed by bootstrap analysis (**Fig. 7**). Variable $\hat{\gamma}$ across dyads, albeit robustly estimated in each dyad, supports the notion that the distribution of total ICIs, which has been shown to obey the power law with an exponent of ~1.3 (**Fig. 2**), is a mixture of power-law distributions with variable exponents.

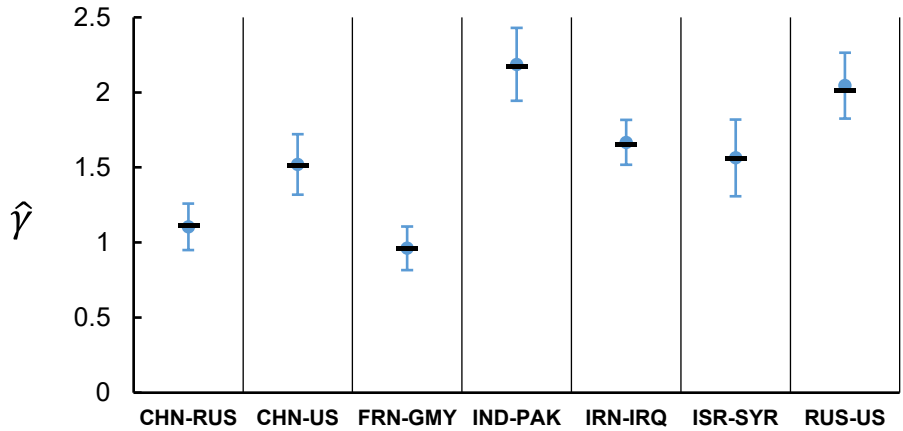

**Figure 7:** The power-law exponent varies from dyad to dyad. The estimated power-law exponent $\hat{\gamma}$ for each dyad is indicated by the filled black bar. To confirm the stability of this estimation, 100 pseudo series of ICIs were synthesized using the bootstrap process, for each of which the power-law exponent was re-estimated. The filled blue circle and error bar indicate the mean and standard deviation of $\hat{\gamma}$ calculated using the bootstrap process.

The mCSN test ensures that the power-law hypothesis is more plausible than the exponential-distribution hypothesis in the estimated domain $\hat{x}_{\min} \leq x \leq x_{\max}$. However, this does not necessarily exclude the possibility that the exponential-distribution hypothesis is more plausible than the power-law hypothesis in another domain. This may have occurred, especially when the number of ICIs was small, as in the present case. To examine this possibility, we conduct an mCSN test to examine the exponential-distribution hypothesis. The results are summarized in **Table S1**. In contrast to the mCSN test of the power-law hypothesis, which gives a $p$-value larger than the criteria of 0.1 for any of the seven dyads, the $p$-value of the mCSN test of the exponential-distribution hypothesis is below the criteria for three dyads (CHN-US, IRN-IRQ, and RUS-US). Therefore, the exponential distribution hypothesis for the estimated domains was excluded from these dyads. The mean difference $\langle \log L^{(\text{p.l.})} \rangle_B - \langle \log \hat{L}^{(\exp)} \rangle_B$ calculated using the bootstrap process was positive for two dyads (CHN-US and FRN-GMY) (**Table S2**), implying that the power-law hypothesis is more likely than the exponential distribution hypothesis in the estimated domains for these dyads. Furthermore, the variability in the estimated $\hat{\lambda}$ appeared to be more sprawling across the dyads (**Fig. S1**) than the estimated $\hat{\gamma}$ (**Fig. 7**), which implies a less robust estimation of $\hat{\lambda}$. Although it is difficult to judge which hypothesis is more plausible, comparing the results shown in **Table 1**, **Table 2**, and **Fig. 7** with those shown in **Table S1**, **Table S2**, and **Fig. S1** strongly suggests that fitting a power-law distribution to the ICI samples for each dyad is more suitable.

**ICIs are independent and identically distributed**

Our information-theoretic model predicts that the ICIs in each dyad are generated independently from an identical power-law distribution. In contrast, the interval $\tau_n$ between the timing of the $(n-1)$-th and $n$-th fatal attacks in insurgency and terrorism approximately follows a power-law progress curve $\tau_n = \tau_1 n^{-b}$, most typically with escalation ($b > 0$) and sometimes with de-escalation ($b < 0$) (Johnson et al. 2011; Johnson et al. 2013). We conducted the following statistical experiment to confirm that the actual generation of ICIs in each dyad was independent and identically distributed and that the observed power-law distribution of ICIs was due to neither escalation nor de-escalation. Let $\tau_n$ be the $n$-th ICI generated in a certain dyad and $a^{(1)}(\boldsymbol{\tau})$ be the first-order autocorrelation calculated for the ICI series $\boldsymbol{\tau} = \{\tau_1, \cdots, \tau_N\}$ (see **Materials and Methods** for details). If series $\boldsymbol{\tau}$ followed escalation or de-escalation, $a^{(1)}(\boldsymbol{\tau})$ would be significantly high. From this series, $B = 10,000$ pseudo series were synthesized by bootstrapping. These pseudo series follow independent and identically distributed processes. We then calculated the distribution of the first-order autocorrelations over these pseudo series. For this distribution, which normally has a single peak around zero, a rejection area is defined rightward with a significance level $p_s$, for which we chose a conservative value ($p_s = 0.1$). This rejection area (red shaded in **Fig. 8**) corresponds to the possibility that the positive correlation between $\tau_n$ and $\tau_{n+1}$ is significantly high, as is the case for escalation and de-escalation. Another rejection area was defined to the left at the same significance level ($p_s = 0.1$). This area (blue shaded in **Fig. 8**), for which the negative correlation between $\tau_n$ and $\tau_{n+1}$ is significantly high, indicates the tendency that a longer ICI is followed by a shorter ICI, and vice versa, thereby producing oscillatory progress.

We can test the null hypothesis that series $\boldsymbol{\tau}$ fails to have nontrivial (i.e., significantly positive or negative) first-order autocorrelation by examining the location of the value for $a^{(1)}(\boldsymbol{\tau})$ in the distribution; if it enters either of the rejection areas, the null hypothesis is ruled out. For any of the seven dyads (CHN-RUS, CHN-US, GMY-FRN, IND-PAK, IRN-IRQ, ISR-SYR, and RUS-US), the value for $a^{(1)}(\boldsymbol{\tau})$, indicated by the vertical lines in **Fig. 8**, is outside the rejection areas. Therefore, the null hypothesis is not rejected for these dyads. It is unlikely that a series without a first-order autocorrelation will have a higher-order autocorrelation. Thus, we concluded that the actual generation of ICIs in each dyad follows an independent and identically distributed process, which is consistent with the predictions of our model.

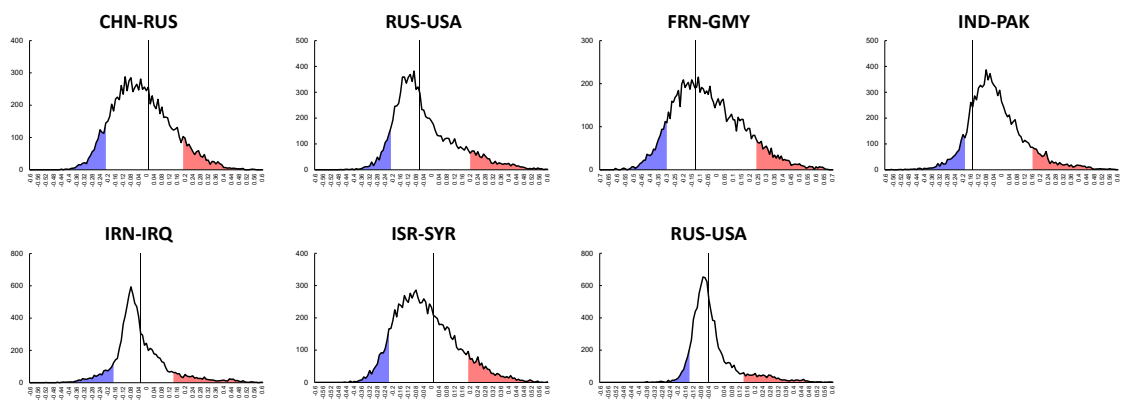

**Figure 8:** ICIs in each dyad are independently generated from an identical distribution. From the actual series of ICIs in each dyad, 10,000 pseudo series were synthesized by bootstrapping. ICIs in each pseudo series conform to independent generation from an identical distribution. The curve in each panel shows the distribution of first-order autocorrelations calculated for the 10,000 pseudo series. The leftward and rightward 10% areas (blue and red shading, respectively) reject the null hypothesis that the actual series of ICIs fails to exhibit nonvanishing first-order autocorrelation. The vertical solid line in each panel indicates the first-order autocorrelation $a^{(1)}(\boldsymbol{\tau})$ of the actual series.

**Discussions**

The severity of war, measured by battle deaths, has been well documented to follow power law since Richardson's proposition in his seminal works (Richardson 1948; Richardson 1960). This study demonstrated that power law also resides in the temporal aspects of interstate conflicts. To this end, we define the inter-conflict interval (ICI) as a critical quantity for exploring temporal statistics. We find that the ICIs compiled from the history of interstate conflicts from 1814 to 2014 follow a power-law distribution. We then propose an information-theoretic model to account for our empirical findings. The model assumes that a pair of states constituting a dyad, struggling with their national interests and survival, will act to balance the principles of promptness and seriousness. The former and latter principles are mathematically formulated as constraints to reduce the generalized mean of the ICI and maximize the mutual information between the timings of consecutive occurrences of conflict. Under these constraints, entropy maximization yields a point process with a point-to-point interval that obeys the power law. The model predicts that a series of ICIs in each dyad are independently generated from an identical power-law distribution. However, the power-law exponent may vary from dyad to dyad. We statistically analyzed individual dyads separately and obtained results consistent with the predictions of the model.

To test the power law hypothesis of ICIs collected from all dyads or collected separately from individual dyads, we used a modified version of the rigorous statistical method proposed by Clauset et al. (2009). This method, which we call the mCSN test, calculates the $p$-value and then judges whether a specific hypothesis (e.g., the power-law hypothesis or the exponential-distribution hypothesis) of ICIs is plausible if the obtained $p$-value is larger than the criteria of 0.1; otherwise, it is ruled out. Noticeably, the $p$-value for ICIs collected from all dyads, plotted as a function of $x_{\max}$ (**Fig. 3a**), shows a conspicuous trough at approximately 9,000 days (~25 years), even though the $p$-value around this trough is slightly larger than 0.1, indicating that the power-law hypothesis is barely plausible. We suppose that this trough was attributable to the interwar period bounded by the end of WWI (1918) and the beginning of WWII (1939). The power-law hypothesis of ICIs premises that the conflict process in each dyad is independent of those in other dyads. However, this premise was apparently violated during WWI and WWII when many countries became involved in war almost simultaneously and automatically, according to either side of the opposing camps they had taken. The resulting excess number of ICIs, whose lengths were comparable to those of the interwar period, eventually caused a substantial deviation in the tail shape of the distribution from the power law. To confirm this supposition, we removed ICIs related to either of the world wars by leaving ICIs whose end and start dates were before the start of 1914 and after the end of 1945, respectively, and then applied the mCSN to the remaining samples. With this prescription, trough levels disappeared (**Supplementary Materials, Fig. S2**). This implies that WWI and WWII, in which interstate wars occurred worldwide and cooperatively, were historically unique events.

Except during WWI and WWII, the results obtained in the present study support the idea that conflict processes in individual dyads are independent of each other. Nevertheless, examining in more detail the influence of conflict processes in some dyads on others, if any, is of substantial interest, as recent studies suggest that higher-order interactions, as well as pairwise (i.e., first-order) interactions between states, affect conflict occurrence (Crammer & Desmarais 2017; Li et al. 2017; Olivella et al. 2022).

The ICI samples collected from all dyads were well-suited to a power-law distribution with an exponent of ~1.3 (**Fig. 3b**). In general, the value of the power-law exponent is related to the frequency of event occurrence; the larger the power-law exponent, the more frequent the events. As the dataset includes interstate conflicts over the past 200 years (1816–2014), a question arises: Is the power-law exponent consistent or changing over the past 200 years? A recent study conducting out-of-sample cross-validation demonstrated that causal models of war vary periodically (Jenke & Gelpi 2017). To address this, we also examined the power-law hypothesis by dividing the entire period (1816–2014) into the following eras: (i) the first half of the 19th century (1816–1858); (ii) the second half of the 19-th century (1859–1899); (iii) the first half of the 20-th century lasting from 1900 to just after WWII (1946); (iv) the Cold War era (1947–1989), and (v) the post-Cold War era (1990~the present (2014)). The results obtained demonstrate that, as time passes, the value of the power-law exponent gradually increases ($\hat{\gamma}$ =1.18, 1.0, 1.22, 1.7, and 1.71 for eras (i), (ii), (iii), (iv), and (v), respectively; see **Supplementary Materials, Table S3,** and **Fig. S3**). A gradual increase in the frequency of conflict over the last 200 years is the most naïve interpretation of these observations. However, the observed increase in the power-law exponent can be attributed to a recording bias. Some interstate conflicts, especially those in older eras, may have been overlooked when compiling the data.

The *p*-value for the second era (iii), lasting from 1900 to 1946, was 0.0286 ($< 0.1$) (**Table S3**). Therefore, the power-law hypothesis is not plausible. As demonstrated in **Fig. S2**, the ruling out of the power law hypothesis was most likely caused by the inclusion of the interwar period in this era. Therefore, we trimmed the last six years of this era. ICIs compiled from this trimmed-off era, lasting from 1900 to 1938, no longer involved an excess number of ICIs compared to the interwar period. Indeed, we obtained $p = 0.449$ ($> 0.1$) for this trimmed-off era, say (iii'), confirming the plausibility of the power-law hypothesis (**Table S3**).

This study demonstrates that the ICI, the interval bounded by consecutive conflicts occurring in the same dyad, follows the power law. In contrast, Richardson's earlier works (Richardson 1944, 1945) suggested that the timing of onset of full-scale wars (interstate wars, in our terminology), occurring anywhere in the world, obeys a Poisson process; that is, the interval between the timing of onset of consecutive wars occurring anywhere in the world follows an exponential distribution. Therefore, we sought to examine whether the timing of onset of interstate conflicts, counted without specifying the dyad, also follows an exponential distribution. To this end, we defined the dyad-unconditioned inter-conflict interval (DUC-ICI, **Fig. S4**). Considering the possibility that the rate of conflict occurrence may change over the years (Clauset 2018), we used the above division of the entire period into five eras. DUC-ICI samples were compiled separately for each era and then underwent the mCSN tests. The results of the mCSN tests for the power-law hypothesis (**Fig. S5**, **Table S4**, and **Table S5**) and the exponential-distribution hypothesis (**Fig. S6**, **Table S6**, and **Table S7**) indicate that the DUC-ICIs for each era is more likely to follow an exponential distribution than a power-law distribution, consistent with the Richardson's suggestion.

We do not ask whether the action taken by either state is a rational means of achieving its political objective, whether the political objective itself is reasonable, or whether it is achieved as intended by settling the conflict. This contrasts with the game-theoretical approach to interstate wars, which assumes that actors behave rationally. This approach has been favored in mainstream international relations theory. For instance, in his game-theoretic model with the assumption that states are 'rational' actors, James Fearon (1995) demonstrated that 'inefficient' (in the sense that they cannot reach a deal that is mutually less costly than an armed confrontation) war can take place between them due to a lack of communication, intentional or unintentional. Our information-theoretic model, which regards armed violence as a means of communication in and of itself, may appear contradictory to Fearon's model as we argue that serious acts do not necessarily conform to rational ones. Nevertheless, we also argue that states only consider the timing of the previous conflict in deciding when to initiate an armed conflict, thereby disregarding its means and costs. Therefore, our insights resonate with the motivation behind the Fearon model.

Our information-theoretic model argues that ICIs are independently generated from an identical power-law distribution in each dyad. The absence of a first-order autocorrelation for a series of actual ICIs in the individual dyads supports this notion. However, the present study did not examine whether autocorrelation resides in the size of interstate conflicts; for instance, every time a conflict occurs, its size grows or shrinks. This issue will be addressed in future studies.

Our model is based on an information-theoretic formulation of the hypothesis that military force is a form of interstate communication. The lines of evidence obtained by the statistical analysis of the MID4.02 dataset support the plausibility of this hypothesis. This hypothesis might contradict the widespread view that interstate war arises from a lack of communication between states. However, from an information-theoretic perspective, the observed power-law property of ICI is the hallmark of maximally efficient communication through violent means.

Power laws are ubiquitously observed in the time course of human behavior, such as e-mail/surface-mail correspondence and web browsing (Barabashi 2005; Vazquez et al. 2006). What makes our findings unique lies in the argument that the power-law property of ICI arises from the interaction between rival states, which we model as a form of communication in the information-theoretic framework. Here, we draw on Clausewitz's statement, which supports this argument.

"War, however, is not the action of a living force upon a lifeless mass … but always the collision of two living forces." (Clausewitz 1823).

Indeed, two classes of models, priority queuing models and modulated Markov processes, have been discussed to account for the power-law property of inter-event intervals empirically observed in human behavior. In contrast to our information-theoretic model, these models assume that individuals behave independently. Priority queuing models (Barabashi 2005; Vazquez et al. 2006; Wu et al. 2010; Vajna et al. 2013) assume that a person has a prioritized list of tasks and executes any task at a time that is probabilistically selected from this list according to this priority. The waiting time of a task from its entry into the list to its execution follows the power law (of exponent 1.0 or 1.5). Each individual creates a prioritized list of tasks that are independent of others. For the modulated Markov process (Malmgren et al. 2008; Malmgren et al. 2009; Karsi et al. 2012; Okada et al. 2021), the power law is accounted for as a consequence of the combination of Poisson processes, which model the behavior of every individual as the sporadic execution of tasks with circadian or weekly cycles. Thus, modulated Markov process models lack the perspective of the interaction between living agents. The above overview of priority queuing and modulated Markov process models suggests that these models cannot explain the observed power-law properties of ICIs, which are thought to be an essential consequence of the interaction between rivalling states. The interactions between rival agents are at the heart of our information theory model. This implies that our information-theoretic model is more favorable than priority-queuing models or modulatory Markov processes for accounting for the power-law properties of ICIs.

A relevant example can be found in a completely different field of neuroscience, where inter-spike intervals (ISIs) in neuronal spike trains have been observed to follow a power law (Kemuriyama et al. 2010; Tsubo et al. 2012). Computational neuroscientists examined the power-law properties of ISIs using the principles of information theory (Tsubo et al. 2012). Neurons, like states, are communicators, and information processing in the brain is the totality of the communication between neurons. Power laws may be a hallmark of communication between real-world actors, such as states or neurons.

This study focuses on armed conflicts between normal states. This contrasts with the recent trend in the discipline, which is devoted to asymmetric warfare, such as insurgency or terrorism, rather than armed conflicts between normal states (Clauset et al. 2007; Bohorquez et al. 2009; Johnson et al. 2011; Clauset and Gleditsch 2012; Johnson et al. 2013; Picoli et al. 2014). The September 11th attacks might have triggered this trend. However, the full-scale war in Ukraine, started by the Russian invasion on February 24, 2022, disenchanted us from the illusion that armed conflict between normal states may be outdated (Kaldor 2013).

## Materials and Methods

### Dataset

This study used the dataset MID4.02, which can be downloaded from a public repository run by the COW Project (https://correlatesofwar.org). The dataset records militarized interstate disputes (MIDs) and interstate wars from 1816 to 2014. Each MID or interstate war in the dataset is specified with a dyad (a pair of states) engaged in the interstate conflict, the start and end days of the conflict, and values for other covariates. For instance, the covariate WAR takes 1 for interstate war and 0 for MID.

### Inter-conflict interval (ICI)

The inter-conflict interval (ICI) is defined as the interval between a conflict in a dyad and the start of the next conflict in the same dyad (**Fig. 1**). Let $t_c^{(\text{start})}$ and $t_c^{(\text{end})}$ be the start and end times of the $c$-th conflict that occurred in a certain dyad, respectively. ICIs were collected from this dyad by calculating $t_{c+1}^{(\text{start})} - t_c^{(\text{end})}$ in ascending order of $c$. If $t_{c+1}^{(\text{start})} - t_c^{(\text{end})} \leq 0$, the $c$-th and $c+1$-th conflicts are regarded as being continued, and therefore, non-positive ICIs are excluded from sampling. In doing so, we obtained 2,369 ICI samples from all dyads. Individual dyads such as CHN-RUS, CHN-US, GMY-FRN, IND-PAK, IRN-IRQ, ISR-SYR, and RUS-US had 38, 29, 20, 33, 29, 33, and 39 ICIs, respectively.

### Goodness-of-fit test for the power-law hypothesis

Clauset et al. (2009) proposed a goodness-of-fit test to examine whether a given set of samples $\{x_1, \cdots, x_N\}$ follows a power-law distribution. This test, which we call the Clause-Shalizi-Newman (CSN) test, was designed to examine the power-law properties of spatial features, such as war size, earthquake magnitude, and urban population. Samples of spatial features, if they follow power-law distributions, include a number of large-sized events because the long tails characterizing power-law distributions imply the likely occurrence of large-sized events. Therefore, the power-law hypothesis to be examined by the original CSN test is mathematically expressed as $p(x) = x^{-\gamma}/\zeta(\gamma, x_{\min})$ $(x_{\min} \leq x)$, where $\sum_{x_{\min}}^{+\infty} x^{-\gamma} = \zeta(\gamma, x_{\min})$ is the generalized zeta function. Note that the domain, $x_{\min} \leq x$, has no upper bound.

In contrast, caution is required when applying the CSN test to temporal features such as ICIs. Sampling ICIs from the dataset MID4.02 is restricted by the recording period used to construct this dataset, which is approximately 200 years, from 1816 to 2014. Therefore, even if ICIs were generable from power-law distributions without upper bounds, the lengths of the ICI samples collected from MID4.02 would never exceed the recording period. This means that the empirical distribution of ICIs has an upper bound, above which no sample exists. Furthermore, individual dyads had their own ages, some of which were much shorter than the recording period. For instance, the Russia-Ukraine dyad was approximately 23 years old in 2014 (the final year of the recording period). The length of the ICI samples collected from the dyads never exceeded their age. Consequently, the empirical distribution of the ICIs collected from all dyads would have an effective upper bound that might be much smaller than the recording period, above which the power-law distribution would no longer fit the data well.

Therefore, to fit a power-law distribution to temporal features, such as ICIs, we must consider the upper bound $x_{\max}$ in addition to the lower bound $x_{\min}$. The power law implies the likely occurrence of long-term events for temporal features. However, such long-term events could not be recorded because the recording period was limited. In contrast, when recording spatial features, large-sized events, such as huge wars (WWI or WWII), huge earthquakes, or megacities, would never be overlooked.

To examine the power-law hypothesis of the ICIs, the original CSN test should be modified by considering the possible presence of upper bounds. The procedure for the modified CSN (mCSN) test, used to examine the power-law hypothesis for ICIs in the present study, was as follows: Let $p(x)$ be the probability distribution of variable $X$. We consider the case where $x$ takes discrete values measured in days. The power-law hypothesis to be examined by the mCSN is mathematically expressed in the following form: $p(x) = x^{-\gamma}/Z(\gamma)$, where $\gamma$ is the power-law exponent and $Z(\gamma) = \sum_{x=x_{\min}}^{\max} x^{-\gamma}$ is the normalization factor equivalent to the partition function. Let $\mathcal{D} = \{x_1, \cdots, x_N\}$ be the data. Samples that are smaller than $x_{\min}$ or larger than $x_{\max}$, if they exist, are excluded from $\mathcal{D}$ because we want to test the hypothesis in the domain $x_{\min} \leq x \leq x_{\max}$. The log-likelihood is then given as

$$L(\gamma) = \sum_{n=1}^{N} \log p(x_n) = -\gamma \sum_{n=1}^{N} \log x_n - N \log Z(\gamma). \tag{12}$$

The value of the power exponent $\gamma$ is determined using the maximum likelihood estimate (MLE). The estimation can be performed by direct numerical maximization of $L(\gamma)$. The model fitted by MLE is denoted as $\mathcal{M}$.

The distance between data $\mathcal{D}$ and the hypothesis is measured by the Kolmogorov-Smirnov (KS) statistic $D_{KS}$ defined by

$$D_{KS} = \max_{x_{min} \leq x \leq x_{max}} |S(x) - P(x)|, \tag{13}$$

where $S(x) = $ (the number of $x_n \geq x)/N$ is the cumulative distribution function (CDF) for the empirical data. $P(x) = \sum_{x \geq x'} p(x')$ is the CDF for the fitted model $\mathcal{M}$.

A large number $S$ of power-law distributed data, $\mathcal{D}_1, \cdots, \mathcal{D}_S$, are synthesized from $\mathcal{M}$. Each data has the same number $N$ of elements as the empirical data $\mathcal{D}$. We fit each synthetic data $\mathcal{D}_s$ to its own power-law model $\mathcal{M}_s$. Then we calculate the KS statistics $D_s$ for $\mathcal{D}_s$ relative to $\mathcal{M}_s$. Then, we count the fraction of time that $D_s$ is larger than $D$, which serves as the $p$-value of this test. Clauset, Shalizi, and Newman (2009) set the conservative decision criteria for the test: If $p \leq 0.1$, the power-law hypothesis for the data $\mathcal{D}$ is ruled out; otherwise, it is plausible. We conducted a goodness-of-fit test for the ICI samples with $S = 10{,}000$ times the generation of synthetic data.

**Mixture of power-law distributions well approximated by a single power-law distribution**

We prove that the likelihood of a mixture of power-law distributions is as close as possible to that of a single power-law distribution. Although the proof is not mathematically rigorous, it provides an intuitive understanding of why a mixture of power-law distributions can be approximated using a single power-law distribution in several cases.

Consider a mixture of power-law distributions:

$$p(x) = \sum_{k=1}^{K} \pi(k) p(x|k), \tag{14}$$

where $p(x|k)$ is the power-law distribution with exponent $\gamma_k$ ($> 1$),

$$p(x|k) = \frac{\gamma_k - 1}{x_{min}} x^{-\gamma_k} \quad (x \geq x_{min}). \tag{15}$$

We assume that the domain of each component distribution has a lower bound $x_{min}$ but infinitely extends rightward without an upper bound. The loglikelihood of the data $\mathcal{D} = \{x_1, \cdots, x_N\}$ is

$$\log L_{mix} = \sum_{n=1}^{N} \log \left( \sum_{k=1}^{K} \pi(k) \frac{\gamma_k - 1}{x_{min}^{-\gamma_k + 1}} x_n^{-\gamma_k} \right). \tag{16}$$

Using Jensen's inequality, one can arrange this as

$$\log L_{mix} \geq \sum_{n=1}^{N} \sum_{k=1}^{K} \pi(k) \log \left( \frac{\gamma_k - 1}{x_{min}^{-\gamma_k + 1}} x_n^{-\gamma_k} \right) = \sum_{n=1}^{N} \left[ \log \frac{1}{x_{min}^{-\gamma+1}} x_n^{-\gamma} + \sum_{k=1}^{K} \pi(k) \log(\gamma_k - 1) \right], \tag{17}$$

where $\gamma \equiv \sum_{k=1}^{K} \pi(k) \gamma_k$. The right-hand side is hence denoted by

$$Q_{\text{mix}} \equiv \sum_{n=1}^{N} \left[ \log \frac{1}{x_{\min}^{-\gamma+1}} x_n^{-\gamma} + \sum_{k=1}^{K} \pi(k) \log(\gamma_k - 1) \right]. \qquad (18)$$

The log-likelihood for a single power-law distribution of the exponent $\gamma$ is given as follows:

$$\begin{aligned}
\log L_{\text{single}} &= \sum_{n=1}^{N} \log \left( \frac{\gamma-1}{x_{\min}^{-\gamma+1}} x_n^{-\gamma} \right) \\
&= \sum_{n=1}^{N} \left[ \log \frac{1}{x_{\min}^{-\gamma+1}} x_n^{-\gamma} + \log \left( \sum_{k=1}^{K} \pi(k)(\gamma_k - 1) \right) \right] \\
&\geq \sum_{n=1}^{N} \left[ \log \frac{1}{x_{\min}^{-\gamma+1}} x_n^{-\gamma} + \sum_{k=1}^{K} \pi(k) \log(\gamma_k - 1) \right] = Q_{\text{mix}}.
\end{aligned} \qquad (19)$$

To derive the inequality in the third row of Eq. (19), Jensen's inequality was used. According to probabilistic machine-learning theories (Bishop 2006), we can solve $\pi(k)$ and $\gamma_k$ by maximizing $Q_{\text{mix}}$. For $\log L_{\text{single}} \geq Q_{\text{mix}}$, an increase in $Q_{\text{mix}}$ leads to an increase in $\log L_{\text{single}}$. Therefore, maximizing $Q_{\text{mix}}$ causes the single power-law distribution with $\gamma = \sum_{k=1}^{K} \pi(k)\gamma_k$ to fit the data more closely. If $\log L_{\text{single}} \geq \log L_{\text{mix}}$, then the single power-law distribution inherently fits the data better than the mixture. Now, consider the case where $\log L_{\text{single}} < \log L_{\text{mix}}$. Since $\log L_{\text{mix}} > \log L_{\text{single}} > Q_{\text{mix}}$, the mixture actually fits the data better than a single power-law distribution. Nevertheless, as $Q_{\text{mix}}$ becomes as close to $\log L_{\text{mix}}$ as possible by its maximization, $\log L_{\text{single}}$, which lies between them, approaches $\log L_{\text{mix}}$. This implies that the mixture can be approximated using a single power-law distribution.

**Testing the power-law hypothesis of ICIs in a single dyad**

The set of 2,369 ICIs collected from all dyads, for which the power-law hypothesis was examined using the mCSN test, was a collection of subsets of ICIs collected from individual dyads. Our information-theoretic model predicts that the power-law hypothesis holds for individual dyads. To test this prediction, we examined whether the ICIs collected from a single dyad followed a power-law distribution. For this purpose, $x_{\max}$ is set to the maximum ICI.

We examined seven dyads (CHN-RUS, CHN-US, GMY-FRN, IND-PAK, IRN-IRQ, ISR-SYR, and RUS-US dyads), each of which provided the number of ICIs eligible for statistical analysis. Nevertheless, the number was relatively low (from 20 to 39 ICIs), which may have caused an overestimation of the $p$-value of the CSN test. Therefore, the obtained $p$-value larger than 0.1, which implies the plausibility of the power-law hypothesis, does not necessarily mean that competing hypotheses, typically the exponential distribution hypothesis, are ruled out. To confirm that the power law hypothesis is more likely than the exponential distribution hypothesis, we compared the log-likelihood between the power law and exponential distribution hypotheses.

Let $\boldsymbol{\tau} = \{\tau_1, \cdots, \tau_N\}$ be the set of ICIs collected from a certain dyad, where $\tau_n$ ($n = 1, \cdots, N$) denotes the $n$-th ICI. We conducted the mCSN test to estimate the lower bound $x_{\min}$ and the power-law exponent $\gamma$, while choosing the upper bound as $x_{\max} = \max_n \tau_n$.

We consider the exponential-distribution hypothesis as a competing hypothesis, which is mathematically expressed as follows: $p(x) = e^{-\lambda x}/Z(\lambda)$ ($x_{\min} \leq x \leq x_{\max}$) with $Z(\lambda) = \sum_{x=x_{\min}}^{x_{\max}} e^{-\lambda x}$. Here, $x_{\min}$ and $x_{\max}$ are the same as those chosen for the power-law fitting. Therefore, let $\hat{\tau}$ be a subset of the ICIs, whose lengths are equal to or greater than $x_{\min}$. Parameter $\lambda$ is estimated by maximizing the log-likelihood:

$$\log L^{(\exp)}(\hat{\tau}; \lambda) = \log \prod_{\tau_n \in \hat{\tau}} p(\tau_n) = -\sum_{\tau_n \in \hat{\tau}} \lambda \tau_n - N \log Z(\lambda), \tag{20a}$$

$$\hat{\lambda} = \arg\max_{\lambda} \log L^{(\exp)}(\hat{\tau}; \lambda). \tag{20b}$$

We compared the maximum log-likelihoods of the exponential distribution given by Eqs. (20) with that for the power-law distribution given by

$$\log L^{(\text{p.l.})}(\hat{\tau}; \gamma) = -\gamma \sum_{\tau_n \in \hat{\tau}} \log \tau_n - N \log Z(\gamma), \tag{21a}$$

$$\hat{\gamma} = \arg\max_{\gamma} \log L^{(\text{p.l.})}(\hat{\tau}; \gamma). \tag{21b}$$

To show that $\log L^{(\text{p.l.})}(\hat{\tau}; \hat{\gamma})$ is significantly larger than $\log L^{(\exp)}(\hat{\tau}; \hat{\lambda})$, we use the bootstrap method to synthesize $B = 100$ pseudo datasets $\hat{\tau}_b$ ($b = 1, \cdots, B$) from $\hat{\tau}$. Then, $\log L^{(\text{p.l.})}(\hat{\tau}_b; \hat{\gamma})$ and $\log L^{(\exp)}(\hat{\tau}_b; \hat{\lambda})$ averaged over the pseudo-datasets were compared by calculating their difference $\Delta^{(l.l.)} = \langle \log \hat{L}^{(\text{p.l.})} \rangle_B - \langle \log \hat{L}^{(\exp)} \rangle_B$, where $\langle \log \hat{L}^{(\text{p.l.})} \rangle_B = \sum_{b=1}^{B} \log L^{(\text{p.l.})}(\hat{\tau}_b; \hat{\gamma})/B$ and $\langle \log \hat{L}^{(\exp)} \rangle_B = \sum_{b=1}^{B} \log L^{(\exp)}(\hat{\tau}_b; \hat{\lambda})/B$. The statistical significance of $\Delta^{(l.l.)} > 0$ was examined using a paired $t$-test.

**Testing independent generation of the ICI series from an identical distribution**

Our information-theoretic model also predicts that the ICI series for each dyad is generated independently from an identical power-law distribution. To validate this, we conducted a statistical test to examine whether the autocorrelation was significantly different from zero. The first-order autocorrelation of $\tau$ is given by

$$a^{(1)}(\tau) = \frac{\sum_{n=1}^{N-1}(\tau_{n+1} - \mu)(\tau_n - \mu)}{(N-1)\sigma^2}, \tag{22}$$

where $\mu \equiv \sum_{n=1}^{N} \tau_n/N$ and $\sigma^2 \equiv \sum_{n=1}^{N}(\tau_n - \mu)^2/N$ are the mean and the variance, respectively (Goh & Barabashi 2008). If the ICIs are independently and identically distributed, the autocorrelation theoretically vanishes. However, as the number of ICIs in each dyad is limited (from 20 to 39 ICIs) in the dyads examined, $a^{(1)}(\tau)$ actually takes either a positive or negative value. Therefore, we tested the null hypothesis that $a^{(1)}(\tau)$ is approximately zero.

To this end, we generated $B = 10{,}000$ pseudo series $\tau_b$ ($b = 1, \cdots, B$) from $\tau$ by bootstrapping. Each pseudo-series satisfied the independent and identically distributed conditions. We then calculated the first-order autocorrelations $a^{(1)}(\tau_b)$ for these pseudo-series and examined their distributions. The 10% left and 10% right areas of this distribution were selected as rejection areas. The null hypothesis is rejected if $a^{(1)}(\tau)$ enters either the left or the right rejection area. If $a^{(1)}(\tau)$ entered the rightward rejection area, it was considered significantly positive. A positive $a^{(1)}(\tau)$ indicates the tendency of ICIs to become progressively longer or shorter. If $a^{(1)}(\tau)$ entered the left rejection area, it was considered significantly negative. A negative $a^{(1)}(\tau)$ implies oscillating ICI series. If $a^{(1)}(\tau)$ enters neither the rightward nor the leftward rejection areas, the null hypothesis cannot be ruled out. It is unlikely that higher-order autocorrelations are significantly positive or negative. whereas first-order autocorrelation vanishes. Therefore, if the above statistical test does not reject the null hypothesis, we conclude that the ICI series is free from autocorrelation, that is, the ICIs are independently generated from an identical distribution.

# Supplementary Materials for

**Testing the Power-Law Hypothesis of Inter-Conflict Intervals**
Hiroshi Okamoto* *et al.*

*Corresponding author. Email: okamoto@coi.t.u-tokyo.ac.jp

**Examining the plausibility of the exponential-distribution hypothesis**

In the main text, we demonstrated that, for any of the seven dyads we examined, the power-law hypothesis is more plausible than the exponential-distribution hypothesis for ICIs equal to or larger than the estimated lower bound $\hat{x}_{\min}$ (**Tables 1**, **2** and **Fig. 7**). However, these results do not necessarily exclude the possibility that the exponential-distribution hypothesis is more plausible in another domain (i.e., for another value $\hat{x}_{\min}$ of the lower bound). To examine this possibility, we conducted an mCSN test of the exponential-distribution hypothesis. The results are shown in **Tables S1**, **S2** and **Fig. S1**. Comparing these results with those shown in **Tables 1**, **2** and **Fig. 7**, we conclude that fitting with a power-law distribution is more suitable. For details, see **Discussion** in the main text.

**The trough in the profile of the $p$-value of the mCSN test is associated with the WWI and WWII**

The profile of the $p$-value as a function of the upper bound $x_{\max}$, given by conducting the mCSN test applied to 2,369 ICIs compiled from all dyads over the entire period (1816~2014), exhibits a conspicuous trough around 9,000 days (~25 years) (**Fig. 3a**). To confirm that this trough is attributed to the WWI and WWII, when the assumption that the process of conflict occurrence in each dyad is independent of those in other dyads is apparently violated, we removed ICIs supposed to be related to either of the world wars and then re-examined the mCSN test applied to the remaining, 2070 ICI samples. Results obtained are shown in **Fig. S2**. The trough disappeared from the profile of the $p$-value (**Fig. S2a**), demonstrating that the trough shown in **Fig. 3a** is actually related to the WWI and WWII. For detailed interpretations, see **Discussion** in the main text.

**Shift of the power-law exponent over the years**

We additionally examined whether the power-law exponent is consistent or changing over the years. To this end, we divided the entire period (1816~2014) into the following eras and then conducted the mCSN test applied to each era: (i) the first half of the 19-th century (1816~1858), (ii) the second half of the 19-th century (1859~1858), (iii) the first half of the 20-th century lasting from 1900 to the end of WWII (1946), (iv) the Cold War era (1947~1989), and (v) the post-Cold War era (1990~the present (2014)). The results shown in **Table S3** and **Fig. S3** demonstrate that the estimated power-law exponent has grown gradually over the last ~200 years. For detailed interpretations of these results, see **Discussion** in the main text.

**The timing of onset of interstate conflicts obeys a Poisson process**

Richardson's earlier works (Richardson 1945, 1946; Clauset 2018) suggested that the timing of onset of full-scale wars (interstate wars, in our terminology), occurring anywhere in the world, conforms to a Poisson process. To examine whether the timing of onset of interstate conflicts, counted without specifying the dyad, also obeys a Poisson process, we defined the dyad-unconditioned inter-conflict intervals (DUC-ICIs, **Fig. S4**). Considering a possible change in the rate of conflict occurrence over the years, we used the above division of the entire period into the following eras under the supposition that the rate of conflict occurrence is constant in each era: (i) 1815–1858, (ii) 1859–1899, (iii) 1900–1946, (iv) 1947–1989, and (v) 1990–2014. DUC-ICI samples compiled for each era underwent the mCSN test of either the power-law hypothesis or the exponential-distribution hypotheses. The results of the CSN test of the power-law hypothesis are summarized in **Table S4**, **Table S5**, and **Fig. S5**. For any of the eras, the $p$-value of the CSN test is below the criteria of 0.1 (**Table S4**) and the power-law hypothesis is less suitable than the exponential-distribution hypothesis in the estimated domain (**Table S5**). The results of the CSN test of the exponential-distribution hypothesis are summarized in **Table S6**, **Table S7**, and **Fig. S6**. For two eras ((ii) and (v)), the $p$-value of the CSN test exceeds the criteria of 0.1, indicating that the exponential-distribution hypothesis is plausible (**Table S6**). For any era, the exponential-distribution hypothesis is more suitable than the power-law hypothesis (**Table S7**). Furthermore, the rate of conflict occurrence, estimated by fitting to either the power-law distribution (**Fig. S5**) or the exponential distribution (**Fig. S6**), grows with time. These results suggest that the DUC-ICI follows is more likely to follow an exponential distribution than a power-law distribution, consistent with the Richardson's suggestion.

**Fig. S1.**

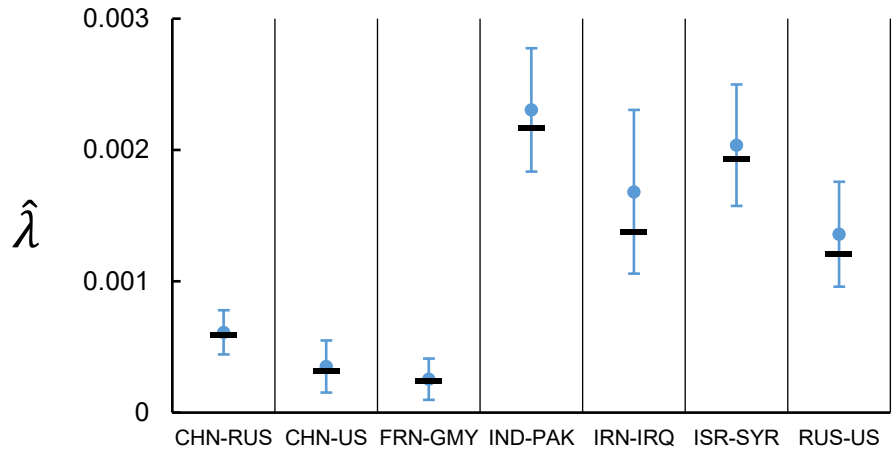

The exponential-distribution parameter $\lambda$ was estimated for each dyad using the mCSN test. The estimated parameter $\hat{\lambda}$ for each dyad is indicated by the black horizontal bar. To examine the stability of this estimation, 100 pseudo series of ICIs are synthesized by bootstrapping, for each of which the parameter value is re-estimated. The filled blue circle and error bar indicate the mean and standard deviation of these values, respectively.

**Fig. S2.**

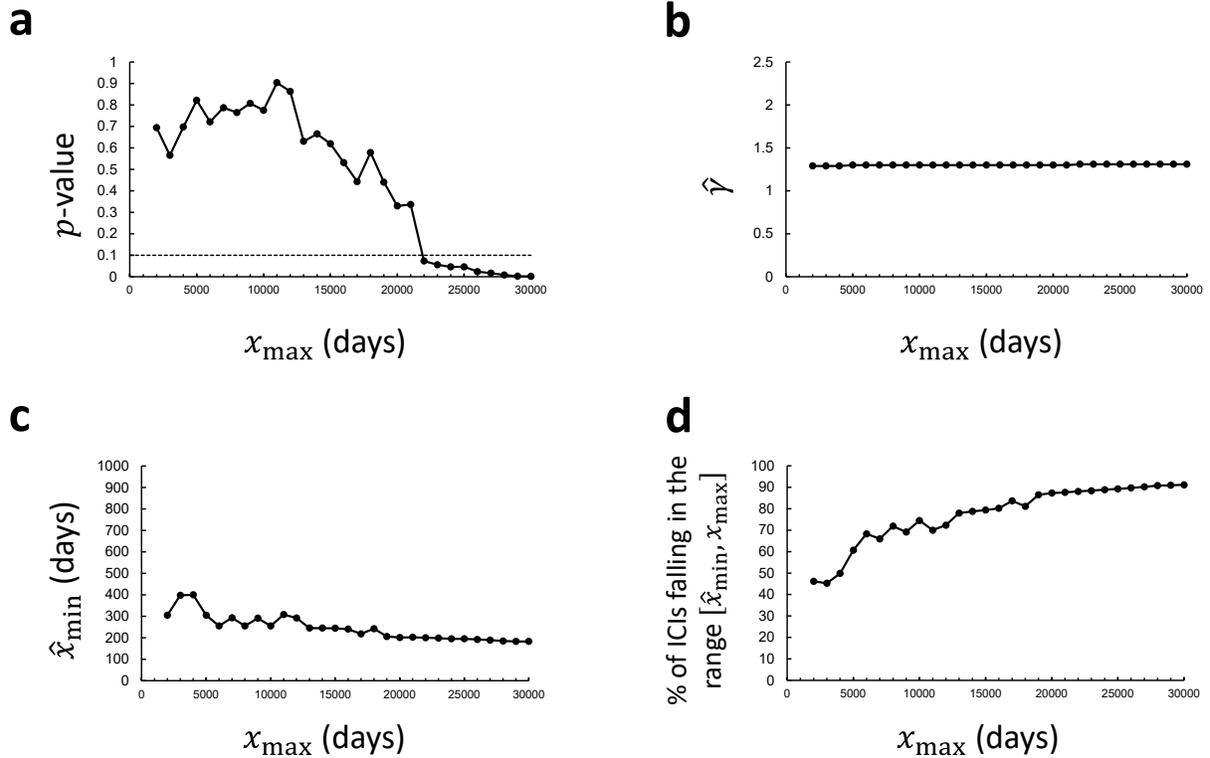

From 2,369 ICI samples collected from all dyads over the entire period (1816~2014), those supposed to be related to the WWI, WWII and interwar period were removed. The same procedures of the CSN test as those for **Fig. 3** were applied to the remaining, 2070 ICI samples. (**a**) The $p$-value of the mCSN test is plotted as a function of $x_{max}$. The horizontal dashed line indicates the criterion of 0.1, for the $p$-value above which the power-law hypothesis is plausible. (**b**) The estimated power-law exponent $\hat{\gamma}$ is plotted as a function of $x_{max}$. (**c**) The estimated lower-bound $\hat{x}_{min}$ is plotted as a function of $x_{max}$. (**d**) The ratio of ICIs (out of the total, 2,369) that fall in the power-law holding domain ($\hat{x}_{min} \leq x \leq x_{max}$) is plotted as a function of $x_{max}$.

**Fig. S3.**

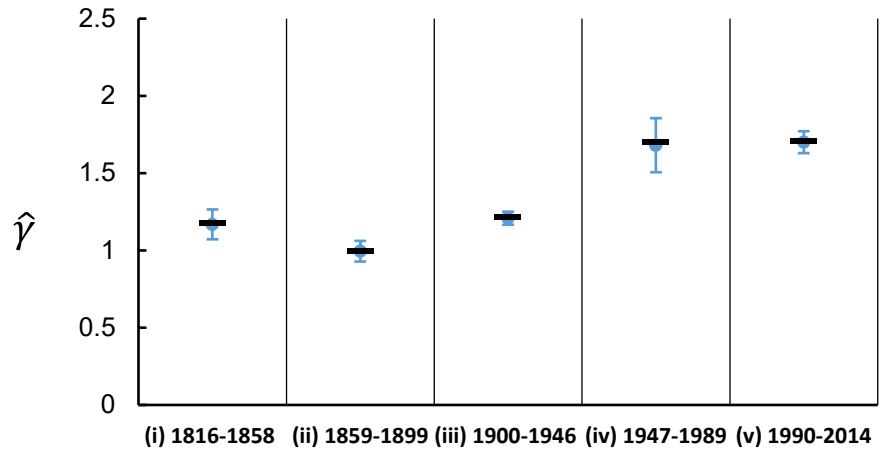

The power-law exponent $\hat{\gamma}$ was estimated for each of the following eras to address whether it is constant or changing over the years: (i) the first half of the 19-th century (1816~1858), (ii) the second half of the 19-th century (1859~1858), (iii) the first half of the 20-th century lasting from 1900 to the end of WWII (1946), (iv) the Cold War era (1947~1989) and (v) the post-Cold War era (1990~the present (2014)). The estimated power-law exponent $\hat{\gamma}$ for each dyad is indicated by the filled black bar. To confirm the stability of this estimation, 100 pseudo-ICI series are synthesized using the bootstrap process, for each of which the power-law exponent is re-estimated. The filled blue circle and error bar indicate the mean and standard deviation of $\hat{\gamma}$.

**Fig. S4.**

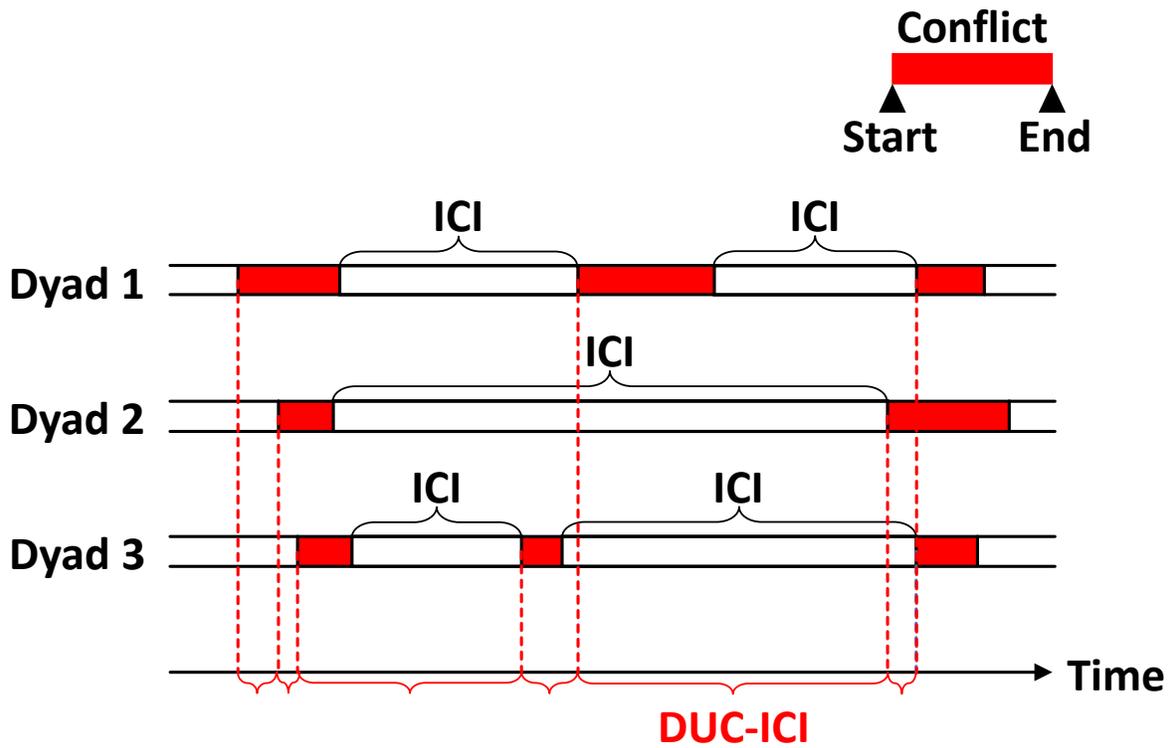

Dyad-unconditioned inter-conflict intervals (DUC-ICIs). The DUC-ICI is the interval between the onset (start) of a conflict and the onset of the next conflict, both counted without specifying the dyad. Each conflict is indicated by the red rectangle.

**Fig. S5.**

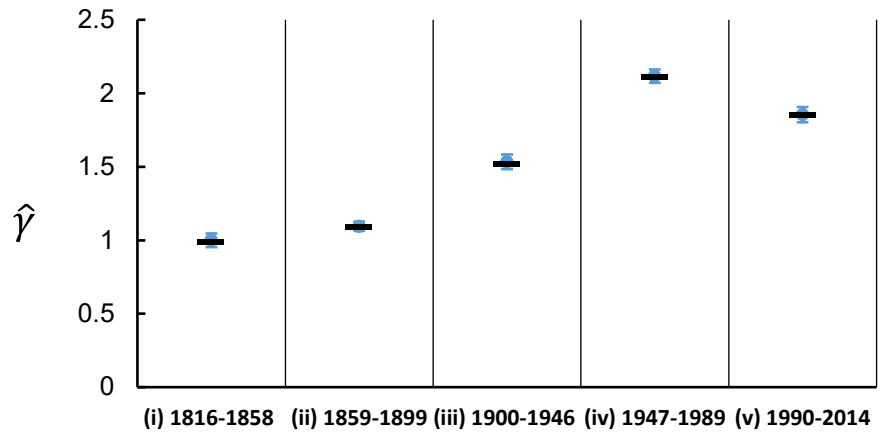

(i) 1816-1858   (ii) 1859-1899   (iii) 1900-1946   (iv) 1947-1989   (v) 1990-2014

The power-law exponent $\hat{\gamma}$ estimated for the following eras: (i) 1815–1858, (ii) 1859–1899, (iii) 1900–1946, (iv) 1947–1989, and (v) 1990–2014. The power-law hypothesis for the DUC-ICIs was examined for each era using the mCSN test. The estimated power-law exponent $\hat{\gamma}$ for each dyad is indicated by the filled black bar. To confirm the stability of this estimation, 100 pseudo-ICI series are synthesized using the bootstrap process, for each of which the power-law exponent is re-estimated. The filled blue circle and error bar indicate the mean and standard deviation of $\hat{\gamma}$.

**Fig. S6.**

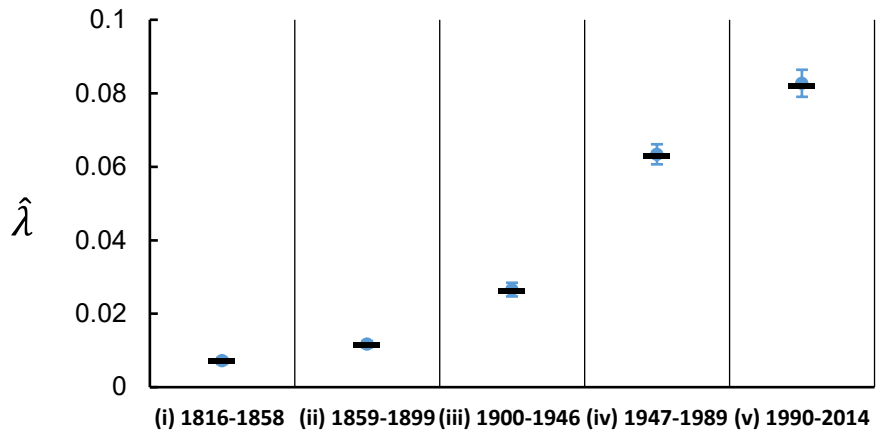

The parameter $\hat{\lambda}$ of the exponential distribution estimated for the following eras: (i) 1815–1858, (ii) 1859–1899, (iii) 1900–1946, (iv) 1947–1989, and (v) 1990–2014. The exponential-distribution hypothesis for the DUC-ICIs was examined for each era using the mCSN test. The estimated parameter $\hat{\lambda}$ for each dyad is indicated by the filled black bar. To confirm the stability of this estimation, 100 pseudo-ICI series are synthesized using the bootstrap process, for each of which the parameter is re-estimated. The filled blue circle and error bar indicate the mean and standard deviation of $\hat{\lambda}$.

**Table S1.**

|  | CHN-RUS | CHN-US | FRN-GMY | IND-PAK | IRN-IRQ | ISR-SYR | RUS-US |
|---|---|---|---|---|---|---|---|
| $N$ | 38 | 29 | 20 | 33 | 29 | 33 | 39 |
| $\hat{x}_{min}$ | 380 | 415 | 379 | 13 | 3 | 3 | 4 |
| $N_D$ | 24 | 14 | 11 | 32 | 29 | 32 | 39 |
| $N_D/N$ | 0.632 | 0.483 | 0.550 | 0.970 | 1.000 | 0.970 | 1.000 |
| $\hat{\lambda}$ | 0.00059 | 0.00032 | 0.00024 | 0.00217 | 0.00138 | 0.00193 | 0.00121 |
| $p$-value | 0.9172* | 0.0012 | 0.5463* | 0.2592* | 0.0039 | 0.176* | 0 |

Results of the mCSN test of the exponential-distribution hypothesis expressed in the form: $p(x) = e^{-\lambda x}/Z(\lambda)$ for $x_{min} \leq x \leq x_{max}$. Here, the value of $x_{max}$ is chosen as the maximum length of ICI samples and the normalization is given by $Z(\lambda) = \sum_{x=x_{min}}^{x_{max}} e^{-\lambda x}$. $N$: the number of ICI samples for each dyad. $\hat{x}_{min}$: the estimated value of $x_{min}$. $N_D$: the number of ICI samples within the domain $\hat{x}_{min} \leq x \leq x_{max}$. $N_D/N$: the ratio of ICI samples within the domain. $\hat{\lambda}$: the estimated value of $\lambda$. The bottom row lists the $p$-value of the mCSN test. For the $p$-value larger than the criteria of 0.1, as indicated by the asterisk (*), the exponential-distribution hypothesis is plausible. The $p$-value below the criteria is colored red.

**Table S2.**

|  | CHN-RUS | CHN-US | FRN-GMY | IND-PAK | IRN-IRQ | ISR-SYR | RUS-US |
|---|---|---|---|---|---|---|---|
| $\langle \log \hat{L}^{(\text{p.l.})} \rangle_B - \langle \log \hat{L}^{(\text{exp})} \rangle_B$ | -0.802 | <span style="color:red">4.391</span> | <span style="color:red">0.141</span> | -11.371 | -10.626 | -1.704 | -12.962 |
| $p$-value | 1.22E-08 | 6.38E-39 | 3.61E-02 | 1.09E-47 | 9.49E-23 | 4.50E-05 | 1.98E-27 |

The upper row lists the mean difference $\langle \log \hat{L}^{(\text{p.l.})} \rangle_B - \langle \log \hat{L}^{(\text{exp})} \rangle_B$ for each dyad. The mean $\langle \log \hat{L}^{(\text{p.l.})} \rangle_B$ was calculated by averaging the loglikelihood for the power-law hypothesis over $B = 100$ pseudo series of ICIs generated using the bootstrap process. The mean $\langle \log \hat{L}^{(\text{exp})} \rangle_B$ of the loglikelihood for the exponential-distribution hypothesis was calculated similarly. Positive values of the quantity $\langle \log \hat{L}^{(\text{p.l.})} \rangle_B - \langle \log \hat{L}^{(\text{exp})} \rangle_B$, colored red, indicate that the exponential-distribution hypothesis is less likely than the power-law hypothesis. The bottom row lists the $p$-value of the paired $t$-test for each dyad to demonstrate the significance of the negativity or positivity of this quantity.

**Table S3.**

|  | (i) 1816-1858 | (ii) 1859-1899 | (iii) 1900-1946 | (iii') 1900-1938 | (iv) 1947-1989 | (v) 1990-2014 |
|---|---|---|---|---|---|---|
| $N$ | 64 | 140 | 400 | 252 | 793 | 619 |
| $\hat{x}_{\min}$ | 191 | 209 | 205 | 294 | 461 | 540 |
| $N_D$ | 60 | 127 | 355 | 207 | 452 | 310 |
| $N_D/N$ | 0.938 | 0.907 | 0.888 | 0.821 | 0.570 | 0.501 |
| $\hat{\gamma}$ | 1.18 | 1 | 1.22 | 1.41 | 1.7 | 1.71 |
| $p$-value | 0.3383* | 0.5603* | 0.0286 | 0.4689* | 0.7965* | 0.2259* |

To address whether the power-law exponent is consistent or changing over the years, we divided the past ~200 years (1816~2014) into the following eras: (i) the first half of the 19-th century (1816~1858), (ii) the second half of the 19-th century (1859~1858), (iii) the first half of the 20-th century lasting from 1900 to the end of WWII (1946), (iv) the Cold War era (1947~1989) and (v) the post-Cold War era (1990~the present (2014)). We then conducted the mCSN test of the power-law hypothesis applied to ICIs collected from each era. We also examined the era, say era (iii'), which is given by trimming the last six years of era (iii) to eliminate the influence of the interwar period. Notations used here are the same as those used for **Table 1**. For the $p$-value of the mCSN test larger than the criteria of 0.1, as indicated by the asterisk (*), the power-law hypothesis is plausible. The $p$-value below the criteria is colored red.

**Table S4.**

|  | (i) 1816-1858 | (ii) 1859-1899 | (iii) 1900-1946 | (iv) 1947-1989 | (v) 1990-2014 |
|---|---|---|---|---|---|
| $N$ | 141 | 269 | 787 | 1300 | 1052 |
| $\hat{x}_{min}$ | 7 | 6 | 10 | 10 | 7 |
| $N_D$ | 102 | 172 | 338 | 532 | 430 |
| $N_D/N$ | 0.723 | 0.639 | 0.429 | 0.409 | 0.408 |
| $\hat{\gamma}$ | 0.99 | 1.09 | 1.52 | 2.11 | 1.85 |
| $p$-value | 0.0001 | 0 | 0.0002 | 0 | 0 |

Results of the mCSN test of the power-law hypothesis for the DUC-ICIs for each of the following eras: (i) 1815–1858, (ii) 1859–1899, (iii) 1900–1946, (iv) 1947–1989, and (v) 1990–2014. Notations used here are the same as those used for **Table 1**. For the $p$-value of the mCSN test larger than the criteria of 0.1, as indicated by the asterisk (*), the power-law hypothesis is plausible. The $p$-value below the criteria is colored red.

**Table S5.**

|  | (i) 1816-1858 | (ii) 1859-1899 | (iii) 1900-1946 | (iv) 1947-1989 | (v) 1990-2014 |
|---|---|---|---|---|---|
| $\langle \log \hat{L}^{(\text{p.l.})} \rangle_B - \langle \log \hat{L}^{(\text{exp})} \rangle_B$ | -7.558 | -25.805 | -0.722 | -25.784 | -29.415 |
| $p$-value | 7.61E-18 | 9.33E-55 | 0.226 | 8.17E-39 | 8.59E-56 |

The upper row lists the mean difference $\langle \log \hat{L}^{(\text{p.l.})} \rangle_B - \langle \log \hat{L}^{(\text{exp})} \rangle_B$ for each era. The bottom row lists the $p$-value of the paired $t$-test for each era to demonstrate the significance of the negativity or positivity of this quantity. Notations and calculations used here are the same as those used for **Table 2**.

**Table S6.**

|  | (i) 1816-1858 | (ii) 1859-1899 | (iii) 1900-1946 | (iv) 1947-1989 | (v) 1990-2014 |
|---|---|---|---|---|---|
| $N$ | 141 | 269 | 787 | 1300 | 1052 |
| $\hat{x}_{min}$ | 6 | 10 | 10 | 9 | 6 |
| $N_D$ | 104 | 157 | 338 | 564 | 468 |
| $N_D/N$ | 0.738 | 0.584 | 0.429 | 0.434 | 0.445 |
| $\hat{\lambda}$ | 0.0071 | 0.0116 | 0.0261 | 0.0628 | 0.082 |
| $p$-value | 0.0158 | 0.5546* | 0 | 0.0871 | 0.7612* |

Results of the mCSN test of the exponential-distribution hypothesis for the DUC-ICIs for each of the following eras: (i) 1815–1858, (ii) 1859–1899, (iii) 1900–1946, (iv) 1947–1989, and (v) 1990–2014. Notations used here are the same as those used for **Table S1**. For the $p$-value of the mCSN test larger than the criteria of 0.1, as indicated by the asterisk (*), the exponential-distribution hypothesis is plausible. The $p$-value below the criteria is colored red.

**Table S7.**

|  | (i) 1816-1858 | (ii) 1859-1899 | (iii) 1900-1946 | (iv) 1947-1989 | (v) 1990-2014 |
|---|---|---|---|---|---|
| $\langle \log \hat{L}^{(\text{p.l.})} \rangle_B - \langle \log \hat{L}^{(\text{exp})} \rangle_B$ | -7.076 | -25.739 | -0.722 | -36.635 | -36.559 |
| $p$-value | 3.52E-16 | 8.69E-60 | 0.226 | 2.52E-53 | 3.47E-65 |

The upper row lists the mean difference $\langle \log \hat{L}^{(\text{p.l.})} \rangle_B - \langle \log \hat{L}^{(\text{exp})} \rangle_B$ for each era. The bottom row lists the $p$-value of the paired $t$-test for each era to demonstrate the significance of the negativity or positivity of this quantity. Notations and calculations used here are the same as those used for **Table S2**.